\documentclass[preprint,preprintnumbers,prd,amsmath,amssymb,nofootinbib,tightenlines]{revtex4}
\usepackage{graphicx}% Include figure files
\usepackage{dcolumn}% Align table columns on decimal point
\usepackage{bm}% bold math

\begin{document}
\preprint{FERMILAB-PUB-09-102-A-T}
\preprint{UK/TP 09-02}

\title{Dark Matter Constraints from a Cosmic Index of Refraction}

\author{Susan Gardner${}^{1,2}$\footnote{Permanent Address}}

\author{David C. Latimer${}^2$}

\affiliation{
${}^1$Center for Particle Astrophysics 
 and Theoretical Physics Department, 
Fermi National Accelerator Laboratory, Batavia, IL 60510}

\affiliation{
${}^2$Department of Physics and Astronomy, University of Kentucky, 
Lexington, KY 40506-0055}

%\date{\today}

\begin{abstract}
The dark-matter candidates of particle physics invariably possess 
electromagnetic interactions, if only via quantum fluctuations. 
Taken en masse, dark matter can thus engender an index of refraction 
which deviates from its vacuum value. Its presence is signaled through 
frequency-dependent effects in the propagation and attenuation of light.
We discuss theoretical constraints on the expansion of the index of 
refraction with frequency, the physical interpretation of the terms, 
and the particular observations needed to isolate its coefficients. 
This, with the advent of new opportunities to view gamma-ray 
bursts at cosmological distance scales, gives us a new probe of
dark matter and a new possibility for its direct detection. 
% restore and tweak
As a first application we use the time delay determined from radio 
afterglow observations of distant gamma-ray bursts to realize a direct limit on 
the electric-charge-to-mass 
ratio of dark matter of 
$\vert \varepsilon \vert/M <  1 \times 10^{-5} \text{ eV}^{-1}$  at 95\% CL. 
\end{abstract}

\maketitle

Some twenty-three percent 
of the Universe's energy budget is 
in dark matter~\cite{concord1,concord2,Komatsu:2008hk}, yet, 
despite its abundance, little is known of its properties. 
A number of methods have been proposed for the 
detection of dark matter. Such studies typically 
rely, for direct searches, on dark-matter--nucleus 
scattering, and, 
for indirect searches, on two-body annihilation of dark-matter to
Standard Model (SM) particles; 
constraints follow
from the nonobservation of the aftermath of particular two-body interactions. 
In contrast, 
we probe dark matter in bulk to infer constraints 
on its particulate nature. 

We search for dark matter by studying the modification of the 
properties of light upon passage through it. One can study either polarization~\cite{svg} 
or propagation effects; we focus here on the latter. The resulting 
constraints are most stringent if dark matter consists of sufficiently low mass particles, 
be they, e.g., warm thermal relics or axion-like particles, that its number density 
greatly exceeds that of ordinary matter. We thus consider dark {\em matter}. 
Matter effects are signaled by dispersive effects in the speed or attenuation of light. 
We study this by introducing an index of refraction $n(\omega,z)$, whose deviation 
from unity is controlled by the light--dark-matter scattering amplitude in the 
forward direction, i.e., the forward Compton amplitude, as well as by the 
angular frequency $\omega$ of the light and the redshift $z$ 
at which the matter is located. A dark-matter particle need not 
have an electric charge to scatter a photon; it need only couple 
to virtual electromagnetically charged particles 
to which the photons can couple. The scattering amplitude 
is related by crossing symmetry to the amplitude for 
dark-matter annihilation into two photons, so that any 
dark-matter model which gives rise to an indirect 
detection signal in the two-photon final-state~\cite{Ullio:2002pj} 
can also drive the index of refraction of light from unity. 
Its real part is associated with the speed of propagation, 
and we search for its deviation from unity by searching 
for frequency-dependent time lags in the arrival of pulses from distant gamma-ray bursts (GRBs).

The limits on the non-observation of frequency-dependent effects in the speed of 
light are severe.  The best limits from terrestrial experiments control 
the variation in the speed of light $c$  with frequency to 
$| \delta c|/c \lesssim 1\times 10^{-8}$~\cite{hall}, 
but the astrophysical limits are much stronger. The arrival time difference 
of pulses from the Crab nebula bound  $|\delta c|/c \lesssim  5\times 10^{-17}$~\cite{crab,schprl}, 
and still better limits come from the study of  GRBs~\cite{AmelinoCamelia:1997gz,schprl}. 
GRBs are bright, violent bursts of high-energy photons lasting on the order of 
thousandths to hundreds of seconds, and their brightness makes them visible at 
cosmological distances. Time delays which are linear in the  photon energy can occur {\em in vacuo} 
in theories of quantum gravity; the special features of GRBs make them particularly 
well-suited to searches for such signatures of Lorentz violation~\cite{AmelinoCamelia:1997gz}. 
The detection  of  photons of up to $\sim31$ GeV in energy from GRB 090510 severely 
constrains this scenario, placing a lower limit on the energy scale at which 
such linear energy dependence occurs to 
$1.2 \text{ E}_\text{Planck} \approx 1.5 \times 10^{19}$ GeV~\cite{abdo}. 
Although vacuum Lorentz violation and light--dark-matter interactions 
can each induce dispersive effects, 
their differing red shift and frequency dependence render them distinct. 

A model-independent analysis of the deviation of the refractive index 
from unity is possible if we assume that the photon energy is small 
compared to the energy threshold required to materialize the 
electromagnetically charged particles to which the dark matter can couple. 
In models of electroweak-symmetry breaking which address the hierarchy problem, 
a dark-matter candidate can emerge as a by-product. In such models 
the inelastic threshold $\omega_\text{th}$  is commensurate with 
the weak scale, or crudely with energies in excess of 
$\mathcal{O}(200 \text{ GeV})$, as the new particles are produced in pairs. 
If the photon energy $\omega$  satisfies the condition  $\omega \ll \omega_{\rm th}$, 
we can apply the techniques of low-energy physics to the analysis of the forward Compton 
amplitude. Under an assumption of Lorentz invariance and other symmetries, 
we expand the forward Compton amplitude in powers of $\omega$ and give a physical 
interpretation to the coefficients of the first few terms as 
$\omega\to 0$~\cite{GGT,GGT2,Hemmert:1997tj}.  In particular, 
the term in ${\cal O}(\omega^0)$  is controlled by the 
dark matter particles' charge and mass, 
weighted by preponderance, irrespective of all other 
considerations save our assumption of Lorentz invariance~\cite{let1}.

The relationship between the 
index of refraction $n(\omega)$ and the 
forward scattering amplitude $f_{\omega}(0)$ for light 
of angular frequency $\omega$ is well-known~\cite{fermi,newton}, where
we relate $f_{\omega}(0)$ 
to the matrix element ${\cal M}$ of quantum field theory 
to connect to particle physics models of dark matter. 
Using standard conventions~\cite{ps}, we determine 
$n(\omega)  = 1 + (\rho/4M^2 \omega^2) \mathcal{M}_r(k, p \to k,p) \,$,
in the matter rest frame~\cite{fermi}, 
so that $p=(M,\mathbf{0})$ and $k=(\omega, \omega \hat{\mathbf{n}})$
with $\rho$  
the mass density of the scatterers and $M$ the particle mass. 
In our analysis we assume $M\gg T(z)$, where $T(z)$ is 
the temperature of the dark matter at the
red shift of the observed gamma-ray burst. 
Since $T(z)$ should be a factor of some $((1+z)/(1 + z_{\rm prod}))^{1-2}$ smaller than
$T(z_{\rm prod})$ at the moment of its production or decoupling,  
our limits are not restricted to cold dark matter 
exclusively. Moreover, even in the latter case, 
the candidate mass can be as light, e.g., as light as $M\sim6\times 10^{-6}$ eV
in the axion model with Bose-Einstein condensation of Ref.~\cite{sikivieyang}. 
Under the assumptions of causality or, more strictly, 
of Lorentz invariance, as well as of charge-conjugation, 
parity, and time-reversal symmetry in the photon--dark-matter interaction, 
we have~\cite{GGT,Hemmert:1997tj}
$\mathcal{M}_r(k, p \to k,p) = f_1 (\omega) 
\boldsymbol{\epsilon'}^{\,\ast} \cdot \boldsymbol{\epsilon}
+ i f_2 (\omega) 
\mathbf{{\cal S}} \cdot \boldsymbol{\epsilon'}^{\,\ast} \times \boldsymbol{\epsilon}$,
where ${\mathbf{\cal S}}$ is the spin operator associated with the dark-matter particle
and $\boldsymbol{\epsilon}$ ($\boldsymbol{\epsilon'}$) 
is the polarization vector associated with 
the photon in its initial (final) state. 
The functions $f_1(\omega)$ and $f_2(\omega)$ are fixed in terms of the dark-matter electric charge
and magnetic moment, respectively, as $\omega\to 0$~\cite{let1,lapidus,brodsky} 
without further assumption 
 --- it does not even matter if the dark-matter particle is composite. 
The amplitude $\mathcal{M}_r(k, p \to k,p)$ is implicitly a $2\times2$ matrix 
in the photon polarization, and its diagonal matrix elements describe dispersion
in propagation and attenuation~\cite{newton}. 
The $f_2(\omega)$ term describes changes in polarization
with propagation, so that we need not consider it further. 
Under analyticity and unitarity, expanding  $f_1(\omega)$  
for  $\omega \ll \omega_{\rm th}$  yields a series in positive powers of 
$\omega^2$  for which the coefficient of every term of  $\mathcal{O}(\omega^2)$  
and higher is positive definite~\cite{GGT,GGT2}. Thus a term in $n(\omega)$  
which is linear in $\omega$, discussed as a signature of Lorentz 
violation~\cite{AmelinoCamelia:1997gz,ellis}, does not appear if 
$\omega <\omega_{\rm th}$ and the medium is unpolarized.  We parametrize 
the forward Compton amplitude as ${\cal M}_r = \sum_{j=0} A_{2j} \omega^{2j}$, 
where $A_0 = -2 \varepsilon^2 e^2$~\cite{let1,ps}  and the dark-matter millicharge 
is $\varepsilon e$. The terms in $\mathcal{O}(\omega^2)$ and higher 
are associated with the polarizabilities of the dark-matter candidate. 

Dispersive effects in light propagation are controlled by the group velocity 
$v_g$, so that the light emitted from a source a distance $l$ away has an 
arrival time of $t(\omega)=l/v_g$. 
For very distant sources we must also take 
the cosmological expansion into account~\cite{jacobpiran}, 
so that as we look back to a light source at redshift $z$, we note that the 
dark-matter density accrues a scale factor of $(1+z)^3$, whereas 
the photon energy is blue shifted by a factor of $1+z$ relative to its present-day 
value $\omega_0$~\cite{jacobpiran}. 
Thus the light arrival time $t(\omega_0,z)$ is 
\begin{eqnarray}
&&\!t(\omega_0, z) = \!\int_0^z \!\frac{\mathrm{d}z'}{H(z')} 
\Bigg( 1 + \frac{\rho_0(1+z')^3}{4 M^2} \Bigg( \frac{-A_0}{ \left((1+z')\omega_0\right)^{2}}  
\nonumber\\
&&+ A_2 + 3 A_4 (1 + z')^2 \omega_0^2 + 
{\cal O}(\omega_0^4) \Bigg) \!\Bigg) 
\label{cosmictlag}
\end{eqnarray}
with the Hubble rate $H(z') = H_0 \sqrt{(1+z')^3\Omega_M + \Omega_\Lambda}$. 
We employ the cosmological parameters 
determined through the combined analysis of WMAP five-year data 
in the $\Lambda$CDM model with 
distance measurements from Type Ia supernovae (SN) and with baryon acoustic oscillation
information from the distribution of galaxies~\cite{Komatsu:2008hk}. 
Thus the Hubble constant
today is $H_0 = 70.5 \pm 1.3 \,\hbox{km\, s}^{-1}\hbox{Mpc}^{-1}$, whereas 
the fraction of the energy density
in matter relative to the critical density today is $\Omega_M = 0.274 \pm 0.015$ and 
the corresponding fraction of the energy density 
in the cosmological constant $\Lambda$ is 
$\Omega_\Lambda = 0.726 \pm 0.015$~\cite{Komatsu:2008hk}. 

We find that the time delay is characterized by powers of 
$\omega^2$ and unknown coefficients $A_{2j}$. 
Different strategies must be employed to determine them. 
The $A_2$ term incurs no frequency-dependent shift in the speed of light, 
so that to infer its presence one needs a distance measure independent of $z$, 
much as in the manner one infers a nonzero cosmological constant 
from Type Ia supernovae data. 
Interestingly, as $A_2>0$ it has the same phenomenological effect as a nonzero 
cosmological constant; the longer arrival time leads to an inferred larger distance scale. 
Cosmologically, though, its effect is very different as it scales with the 
dark-matter density; it acts as grey dust. The remaining terms can be 
constrained by comparing arrival times for differing observed  $\omega_0$. 

The determination of $A_0$ and $A_4$ require the analysis of the GRB light curves
at extremely low and high energies, respectively, and probe
disjoint dark-matter models. 
As a first application of our method, we
use radio afterglow data to determine $A_0$ and thus to yield 
a direct limit on the electric-charge to mass of dark matter. 
This quantity gives insight into the mechanism of dark
matter stability. 
If dark matter possesses an internal symmetry, e.g., 
it cannot decay to lighter particles and conserve its hidden charge. 
Such dynamics can also conspire to give dark matter a slight 
electric charge~\cite{mirror,feldman,kusenko}, which, no matter how small, 
reveals the existence of its hidden interactions and the reason for its stability. 

\begin{figure}
\includegraphics[scale=0.4]{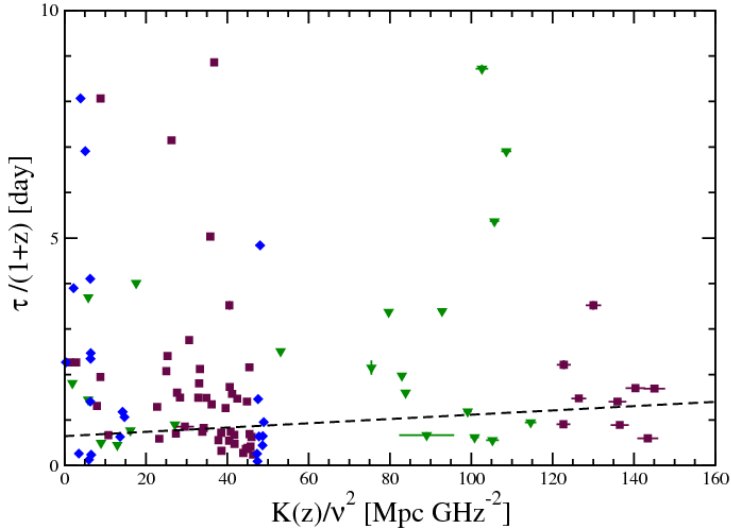}
\caption{The time lag $\tau$ determined from the
observation of a GRB and its radio afterglow 
plotted as a function of $K(z)/\nu^2$ with 
$\nu=\omega/2\pi$, employing the data reported in the supplementary material~\cite{supp}. 
The points correspond to frequency windows 
of 
$4.0- 12$ GHz ($\blacktriangledown$, green), 
$12-30$ GHz ($\blacksquare$, maroon), and 
$30-75$ GHz ($\blacklozenge$, blue) in the GRB rest frame. 
Points with $(1+z)\nu < 4.0$ GHz do not appear within the chosen frame of the figure. 
The fit of Eq.~(\ref{lag}) to the data with  $(1+z) \nu > 4.0$ GHz with a scale factor in the uncertainty in $\tau/(1+z)$ of 450, to compensate for environmental effects in the vicinity of the emission from the GRB, yields $\tilde A_0 = 0.0010 \pm 0.0019$ day Ghz$^2$ Mpc$^{-1}$  
and $\delta = 0.65 \pm 0.10$ day with $\chi^2/\text{ndf} = 1.13$. 
Thus $\tilde A_0 < 0.005$ day Ghz$^2$ Mpc$^{-1}$ at 95\% CL to yield
$\vert \varepsilon \vert/M <  1 \times 10^{-5} \text{ eV}^{-1}$ at 95\% CL.
The statistical scale factors are not shown explicitly. 
For clarity of presentation 
we display time lags in the GRB rest frame of less than 10 days only. } 
\label{fig1} 
\end{figure}

    To determine $A_0$ 
we consider GRBs with known redshift in which a radio afterglow 
is also detected. 
We collect the data and describe the criteria used in its selection in the supplementary 
material~\cite{supp}. 
The time lag between the initial detection of the GRB at some energy and the detection of the radio afterglow 
is $\tau = t(\omega_0^{\rm low},z) - t(\omega_0^{\rm high}, z)$. If we first observe the GRB at keV energies and 
compare with the observed arrival time in radio frequencies, then the terms in positive powers of $\omega_0$, 
as well as the term in $1/(\omega_0^{\rm high})^2$, are negligible; we let  $\omega  \equiv \omega_0^{\rm low}$. 
In order to assess reliable limits on $A_0$ we must separate propagation effects from intrinsic source effects. 
Statistically, we expect time delays intrinsic to the source to be independent of  $z$, and the time delay from 
propagation to depend on  $z$ and $\omega$  in a definite way. Such notions have been previously employed in 
searches for Lorentz invariance violation~\cite{ellis}. We separate propagation and emission effects, respectively, via 
\begin{equation}
\frac{\tau}{1+z} = \widetilde{A}_0 \frac{K(z)}{\nu^2} + \delta((1+z)\nu) \label{lag}
\end{equation}
where   $K(z) \equiv (1+z)^{-1} \int_0^z \mathrm{d}z\,(1+z') H(z')^{-1}$
depends on the cosmological past through the Hubble rate $H(z)$ and $ \delta((1+z)\nu)$  allows for a frequency-dependent time lag for emission from the GRB in the GRB rest frame. The frequency  $\nu \equiv \omega/2\pi$, and  $\widetilde{A}_0$ contains the millicharge-to-mass ratio $\varepsilon/M$, i.e.,    $4\pi^2 \widetilde{A}_0 = - A_0 \rho_0/4 M^2 = 
2\pi\alpha \varepsilon^2 \rho_0/M^2$ with $\rho_0\simeq 1.19\times 10^{-6}$ GeV/cm${}^3$~\cite{Komatsu:2008hk} and $\alpha$ the fine-structure constant. To provide a context, we first consider the value of $|\varepsilon | /M$  which would result were we to attribute the time lag associated with the radio afterglow of one GRB to a propagation effect.  Choosing the GRB with the largest value of $K(z)/\nu^2$, we have a time lag of  $2.700 \pm 0.006$ day associated with GRB 980703A at $z=0.967 \pm 0.001$  measured at a frequency of  1.43 GHz. With Eq.~(\ref{lag}), setting $\delta=0$, 
and noting that  $K(z)/\nu^2 = 1170 \pm 10 \,\hbox{Mpc GHz}^{-2}$  if the errors in its inputs are uncorrelated, the measured time lag fixes  $|\varepsilon|/M\simeq 9 
\times 10^{-6}\,\text{ eV}^{-1}$. Since there are no known examples of a radio afterglow preceding a GRB, this single time lag in itself represents a conservative limit. 
Turning to our data sample of 53 GRBs, we plot the measured time lag versus $K(z)/\nu^2$ in Fig. \ref{fig1} and make a least-squares fit of Eq. (\ref{lag}) 
to determine $\widetilde{A}_0$  and $\delta((1+z)\nu)$. 
We require  $\widetilde{A}_0 > 0$ as demanded by our model. Fitting to the points with frequencies of  $4.0-75$ GHz in the GRB rest frame, we determine  
$\vert \varepsilon \vert/M <  1 \times 10^{-5} \text{ eV}^{-1}$  at 95\% CL, 
which is comparable to our limit derived from a single observation of GRB 980703A.  The dependence of our fit results on the selected frequency window, as well as the stability of our fits to the significance of the radio afterglow observation, to evolution effects in $z$ , and to the more poorly determined red shifts and radio afterglows is discussed in the 
supplementary information~\cite{supp}.

\begin{figure}
\includegraphics[scale=0.4]{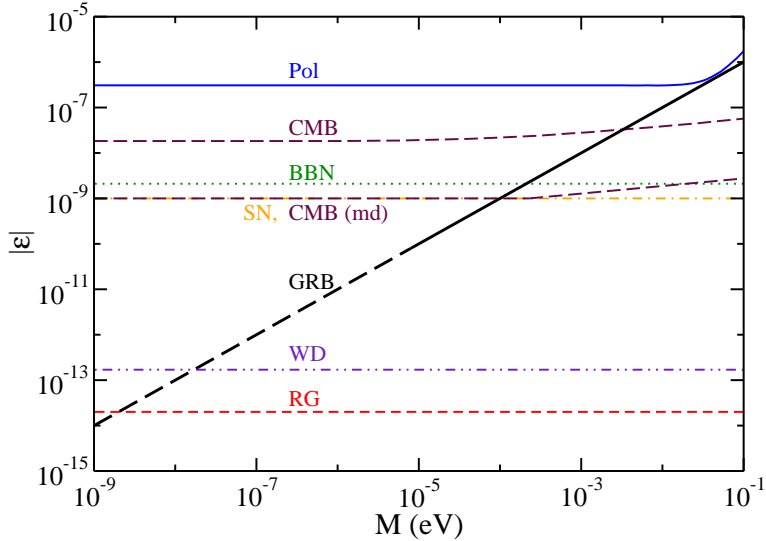}
\caption{Comparison of our direct limit 
on the absolute electric charge of dark matter, $|\varepsilon|$, 
in units of $e$ with candidate mass $M$ with limits from other sources stemming
from millicharged particle production. Our limit 
derives from observations of GRBs and their radio afterglows and is marked 
``GRB'' (solid, black), so that the region above that line is excluded. 
The condition $\omega < \omega_{\rm th}$, as discussed in text, 
sets the lower endpoint of the solid line. Since Eq.~(\ref{lag}) persists in QED, for which
$\omega_{\rm th}=0$, we expect our limit to persist for lighter, cold dark matter as well, as
indicated by the dashed line. The strongest laboratory limits, which are for fermions, are marked 
``Pol'' (solid, blue)~\cite{Ahlers:2007qf}, and the strongest 
limits from induced distortions in the
CMB, which are also for fermions, 
are marked ``CMB'' (long-dashed, maroon) --- the upper curve is the model-independent limit, 
whereas the lower curve is the model-dependent (md) limit~\cite{Melchiorri:2007sq}. 
Constraints on $|\varepsilon|$ emerge from limits on novel energy-loss mechanisms in stars
and supernovae; such limits also fail to act if $|\varepsilon|$ is too large. 
The limit from plasmon 
decay in red-giants is marked by ``RG'' (short-dashed, red)~\cite{davidson}, 
the same limit in white dwarfs
is marked by ``WD'' (dot-dot-dashed, indigo)~\cite{davidson}, 
and the limit from SN 1987A is marked 
by ``SN'' (dot-dashed, orange)~\cite{davidson}. 
%The SN limit acts if $|\varepsilon| \lesssim 10^-7$, whereas 
The RG limit acts if $|\varepsilon| \lesssim 10^{-8}$~\cite{davidson}. 
We have also reported the limit from big-bang nucleosynthesis, marked by 
``BBN'' (dotted, green), from Ref.~\cite{davidson} as well. 
} 
\label{fig2} 
\end{figure}

We have found a direct observational limit on the dark-matter 
electric-charge-to-mass ratio. Our study probes for a charge imbalance 
averaged 
over cosmological distance scales, without regard to its sign, 
at distance scales shorter than the wavelengths of the radio observations in our data set. 
% add charge imbalance? - no
Our bound rules out the possibility of charged ``Q-balls''~\cite{kusenko,shoemaker} 
of less than 100 keV in mass as dark-matter candidates. 
Our limit holds regardless of the manner in which the dark matter is produced, 
though we can compare it to limits arising from the 
nonobservation of the effects of millicharged particle production. 
For example, for $M\sim 0.05\,\text{eV}$, they are crudely comparable to the strongest bound from 
laboratory experiments~\cite{davidson,Ahlers:2007qf}, 
for $|\varepsilon| < 3-4 \times 10^{-7}$ 
for $M\lesssim 0.05\,\text{eV}$~\cite{Ahlers:2007qf}. In comparison the model-independent 
bound arising from induced distortions in the cosmic microwave background (CMB) radiation is 
$|\varepsilon| \le 10^{-7}$  for  $M < 0.1\,\text{eV}$, though model-dependent constraints 
reach $|\varepsilon| \le 10^{-9}$  
for  $M <  2 \times 10^{-4}\,\text{eV}$~\cite{Melchiorri:2007sq}. 
Cosmological limits also arise from observations of the Sunyaev-Zeldovich effect for 
which, e.g., $|\varepsilon| \le 3\times 10^{-7}$  for  
$M \sim 10^{-6}\,\text{eV}$~\cite{burrage}. 
For these light masses our limit is stronger, which shows that millicharged particles 
of such mass and charge are not the primary constituents of dark matter.  Limits also arise 
from stellar evolution and big-bang nucleosynthesis constraints, for which the strongest is 
$|\varepsilon| < 2 \times 10^{-14}$ for $M < 5\,\text{keV}$~\cite{davidson}, as well as 
from the manner in which numerical simulations of galactic 
structure confront observations~\cite{Gradwohl:1992ue,acker,feng}. We offer a visual
summary of this discussion in Fig.~\ref{fig2}. Indirect limits can be evaded: 
for example, in some models, the dynamics which gives rise to millicharged matter 
are not operative at stellar temperatures~\cite{Ahlers:2007qf,massoredondo}; 
other models evade the galactic structure constraints \cite{kusenko,shoemaker}. 

Our limit also significantly restricts 
the phase space of models with hyperweak gauge interactions and 
millicharged particles which can arise in string theory scenarios~\cite{burgess,burrage} as viable 
dark-matter candidates. We estimate that our limit can be improved considerably before the dispersive 
effects from ordinary charged matter become appreciable~\cite{Bombelli:2004tq}. The largest such 
contribution to $\widetilde{A}_0$ should come from free electrons. We estimate the cosmological 
free electron energy density $\rho_e$ to be no larger than 
$\rho_e = (M_e/M_p)\rho_\text{cr} \Omega_b \approx 0.130 \,\text{eV/cm}{}^3$~\cite{Komatsu:2008hk}, 
where $\Omega_b$  is the fraction of the energy density in baryons with respect to the critical density 
today and $M_e$  and $M_p$  are the electron and proton mass, respectively. 
Replacing $\rho_0$ with $\rho_e$  and $\varepsilon/M$  with $1/M_e$  in $ \widetilde{A}_0$ 
we find that our limit would have to improve by $\mathcal{O}(2\times 10^{-3})$ before the contribution 
from free electrons could be apparent. We set our limit of 
$\vert \varepsilon \vert/M <  1 \times 10^{-5} \text{ eV}^{-1}$ from existing radio observations 
at no less than 4 GHz in the GRB rest frame, so that our limit is
certainly operative 
if  $\omega_{\rm th}/2\pi > 4$ GHz, or crudely, if $M>8 \times 10^{-6}$ eV. 
Studies of the polarizability in QED~\cite{llanta}, 
for which $\omega_{\rm th}=0$, also reveal 
the analytic structure in $\omega$ we have assumed for the forward Compton amplitude. 
Thus we believe our limit to be of broader validity, so that 
the lower limit on the mass can be less than $8 \times 10^{-6}$ eV, though it 
is model dependent and set by the Lee-Weinberg constraint~\cite{leew}, 
much as the minimum mass 
of $\sim6\times 10^{-6}$ eV is determined in the axion model of Ref.~\cite{sikivieyang}. 
Forward scattering is coherent irrespective of 
whether the photon wavelength is large compared to the 
interparticle spacing, so that we expect our results to persist in the dilute particle
limit
%, i.e., as $M$ increases, 
as well, as supported by laboratory studies~\cite{campbell}. 
%Even with $M=10^{-6}\,\hbox{eV}$, the cosmological dark-matter number density is 
%still smaller than the number density of the dilute terrestrial 
%system studied in Ref.~\cite{campbell}. 
One further comment: 
at a frequency of 4 GHz our limit implies that we probe the
average net charge of dark matter, with no constraint on its sign, 
at length scales of no longer than 8 cm. 
Our limits can be significantly bettered through GRB radio afterglow studies at longer wavelengths.
%; this also allows to examine the net charge on longer length scales. 

We thank Keith Olive for an inspiring question and Scott Dodelson, Ren\'ee Fatemi, Wolfgang Korsch, 
and Tom Troland for helpful comments. SG would also like to thank Stan Brodsky for 
imparting an appreciation of the low-energy theorems in Compton scattering and 
the Institute for Nuclear Theory and the Center for Particle Astrophysics 
and Theoretical Physics at Fermilab for gracious hospitality. 
This work is supported, in part, by the U.S. Department of Energy
under contract DE-FG02-96ER40989.

%include: ra_table_29Jan.tex
\pagebreak

\begin{center}
\Large{\bf Supplementary Information}
\end{center}

\section{Gamma-Ray Burst Data}

From the publicly available data, we consider the 
gamma-ray bursts (GRB) 
with known redshifts and detected radio afterglows through March, 2009. 
Tables 1--4 collect the GRBs with these properties 
which satisfy certain criteria. We demand that the energy of the initially detected 
gamma-ray burst in its rest frame be compatible with the energy range of the Fermi Gamma-Burst Monitor. This excludes the X-ray flashes 
GRB 080109A and GRB 020903A. To determine whether an observation of the radio flux is indeed a detected radio afterglow, we demand that
the radio flux detection be within the error box of the location of the observed GRB, that it not be a site of previously observed 
radio activity, and that it be significantly non-zero. We use the criterion employed by 
Ref.~\cite{grb080810A_ra} 
and thus require that the radio observation
be nonzero by three standard deviations or more to be termed a detection. Such considerations exclude GRB 030277A, GRB 980425A, and GRB 011130A. 
Consequently, we find 53 GRBs to consider, 
for which we report all detected radio frequencies of 75 GHz or less. 
We report the time of the initial detection of the GRB and the time of the detection of a particular observation frequency $\nu$ in 
the associated radio afterglow (RA). We calculate the time lag between the detection of a particular radio frequency 
and the initial detection of the GRB from these quantities.

Remarks concerning the collected data are in order. 
A redshift marked with an \mbox{asterisk $^*$} means that no certainty has been reported in the work which determines it. In those cases, 
we assume the uncertainty in the redshift to be plus or minus one unit in the last significant figure reported.  
The redshift we report for GRB 021004A is the average of those reported in Refs.~\cite{grb021004A_z1,grb021004A_z2}.
For the redshifts marked with a \mbox{circle $^\circ$}, the associated publication states that the reported redshift is either a lower limit or 
the ``likely'' value.  In the particular case of GRB 970508A, 
work in refinement of the upper bound on the red shift has been reported
in Ref.~\cite{grb970508A_z3}.
For the frequency marked with a double \mbox{dagger $^\ddag$}, 
observations were made in the frequency range from 14.5 GHz to 17.5 GHz.  

We also note which radio detections have lesser significance, so that 
detections with a significance between 3$\sigma$ and 4$\sigma$ are marked with a \mbox{sharp $^\sharp$}, whereas 
detections with a significance between 4$\sigma$ and 5$\sigma$ are marked with a \mbox{flat $^\flat$}.  
For time lags marked with a \mbox{dagger $^\dagger$}, we assume the uncertainty in the time lag to be plus 
or minus one unit in the last significant figure of the reported detection time of a particular radio frequency. 
If a radio frequency detection reports an observation time interval, the time lag is calculated from the 
midpoint of the time interval, and 
we take one half of the observation time as the uncertainty in the time lag.  
The RA time reported in these cases is that of the midpoint of the observation time.

To realize the limits reported in the main body of our paper, 
we fit to Eq.~(2) of that source. 
We find that large statistical scale factors must be employed to realize good fits to the data. 
This may stem, in part, from the circumburst environment, and, more generally, from 
time delay effects which arise from neither source effects 
nor propagation effects across the expanse of space. 
The nature of the circumburst environment of GRB 050904A is discussed in 
Ref.~\cite{grb050904A_ra}.

To study how our limits rely on the details of our data set,  
we compare fits with frequency in the source rest 
frame. We also study how the fits change with the significance of the 
included radio observations, 
with omitting radio observations with reported
integration times, and with omitting more poorly determined red shifts.  
We also try to study source evolution effects with $z$. 
As to the frequency dependence, differences do emerge if we include 
points at the observed frequency of $1.43$ GHz. This portion of the data 
set admits time delays in the GRB rest frame of a month and more, and larger
scale factors are required to yield fits of comparable quality. 
Fitting data satisfying $(1+z)\nu > 4.0$ GHz, for which 
with a scale factor of 450 applied to the error in $\tau/(1+z)$
we find 
$\chi^2/\text{ndf} = 1.13$ 
with 
$\tilde A_0 = 0.0010 \pm 0.002$ day Ghz$^2$ Mpc$^{-1}$  
and $\delta((1+z) \nu) = 0.65 \pm 0.10$ day, 
or 
$\tilde A_0 < 0.005$ day Ghz$^2$ Mpc$^{-1}$  at 95\% CL. 
Were we to fit all the data
points we would require a scale factor of 685 to yield a fit of
$\chi^2/\text{ndf} = 1.13$ from which we would find 
$\tilde A_0 < 0.007$ day Ghz$^2$ Mpc$^{-1}$  at 95\% CL. 
We thus restrict our discussion henceforth to the portion 
of the data set which satisfies $(1+z)\nu > 4.0$ GHz. 
If we now restrict 
our fit to points in which the significance of the radio observation is
$5\sigma$ or more, then our limit on the slope is still 
$\tilde A_0 < 0.005$ day Ghz$^2$ Mpc$^{-1}$ at 95\% CL, 
though we must
increase the scale factor to 512 to recover a fit of comparable quality. 
Some of the RA 
measurements have reported integration times 
and hence determinable uncertainties, whereas others do not. Repeating 
the fits without the points with reported integration times yields no
significant difference in the fit, though we must increase the scale 
factor to 467 to yield a fit of comparable quality. 
This may well follow as the dropped points do 
have larger errors and hence play a lesser role in the original fit. 
If we now return to 
our original data set and omit those points for which the determined
redshift is a lower limit or merely ``likely,'' the limits do weaken. 
For example, with a scale factor of 419, we find $\chi^2/\text{ndf}= 1.13$ 
with $\tilde A_0 = 0.0053 \pm 0.0048$ day Ghz$^2$ Mpc$^{-1}$ 
and $\delta((1+z)\nu) = 0.41 \pm 0.21$,
so that the limit on the slope at 95\% CL is roughly a factor of 3  worse. 
This can be rationalized as these poorly known redshifts possess 
large $z$. Finally, to study the effects of cosmological evolution on the GRBs, we repeat
our fit with $z > 1$ and find similar results. That is, for a scale factor of 539
we determine $\chi^2/\text{ndf}= 1.14$  with 
$\tilde A_0 = 0.0017 \pm 0.0023$ day Ghz$^2$ Mpc$^{-1}$ 
and $\delta((1+z)\nu) = 0.58 \pm 0.12$, to yield 
$\tilde A_0 < 0.006$ day Ghz$^2$ Mpc$^{-1}$ at 95\% CL. 
Thus the effects of cosmological evolution appear to be small. 

\bigskip
\pagebreak

\begin{table}[hp]
\begin{center}
\begin{tabular}{|l|l|l|l|l|l|}
\hline
GRB	&	Redshift		&	GRB Time		&	RA Time	&	Frequency  &	Time Lag \\ 
	&				&		(UT)		& 	(UT)		&  (GHz)     &  (d) \\ \hline
090328A & 0.736 $\pm$ 0.001$^*$ \cite{grb090328A_z} & 09:36.46  \cite{grb090328A_det} & Mar 30.99  \cite{grb090328A_ra} & 8.46 & 2.59 $\pm$ 0.01$^\dagger$ \\ \hline

090323A	& 3.6 $\pm$ 0.1$^*$ \cite{grb090323A_z}  &   00:02:42.63  \cite{grb090323A_det}  & Mar 27.38   \cite{grb090323A_ra} & 8.46 &  4.38 $\pm$ 0.01$^\dagger$ \\ \hline

090313A &  3.375 $\pm$ 0.001$^*$  \cite{grb090313A_z}&  09:06:27   \cite{grb090313A_det}&  Mar 16.13  \cite{grb090313A_ra} & 16$^\ddag$ & 2.75 $\pm$ 0.05 \\ \hline

081007A	&	0.5295 $\pm$ 0.0001 \cite{grb081007A_z} & 05:23:52 \cite{grb081007A_det}	& Oct 09.19 \cite{grb081007A_ra}&	8.46	&1.97 $\pm$ 0.01$^\dagger$ \\ \hline

080810A	&	3.35 $\pm$ 0.01$^*$ \cite{grb080810A_z}	&	13:10:12	\cite{grb080810A_det}	& Aug 13.36 \cite{grb080810A_ra}&	8.46$^\sharp$	&	2.81 $\pm$ 0.01$^\dagger$ \\ \hline

080603A	&	1.6880 $\pm$ 0.0001$^*$	\cite{grb080603A_z}	&	11:18:11	\cite{grb080603A_det}	& Jun 07.42 	\cite{grb080603A_ra}&	8.46	&	3.95 $\pm$ 0.01$^\dagger$ \\ \hline

080319B	&	0.937 $\pm$ 0.001$^*$ \cite{grb080319B_z}	&	06:12:47	\cite{grb080319B_det}	& Mar 21.56	\cite{grb080319B_ra}&	4.86$^\flat$	&	2.30 $\pm$ 0.01$^\dagger$ \\ \hline

%not a GRB
%080109A	&	0.0069 $\pm$ 0.0001$^*$ \cite{grb080109A_z}	& 13:33:53 \cite{grb080109A_det}	&Jan 13.3  \cite{grb080109A_ra}&	8.46	&	3.7 $\pm$ 0.1$^\dagger$	 \\ \hline

071122A	&	1.14 $\pm$ 0.01$^*$ \cite{grb071122A_z}	& 01:23:25 \cite{grb071122A_det}	&Nov 24.94	\cite{grb071122A_ra}&	8.46	&	2.8779 $\pm$ 0.0001$^\dagger$ \\ \hline

071020A	&	2.145 $\pm$ 0.001$^{* \circ}$	\cite{grb071020A_z}	& 07:02:26 \cite{grb071020A_det}	& Oct 22.4671	\cite{grb071020A_ra}&	8.46	&	2.1737 $\pm$ 0.0004$^\dagger$ \\ \hline 

071010B	&	0.947 $\pm$ 0.001$^*$	\cite{grb071010B_z}	& 20:45:47 \cite{grb071010B_det}	& Oct 13.7675	\cite{grb071010B_ra}&	8.46	&	2.9024 $\pm$ 0.0004$^\dagger$ \\ \hline

071003A	&	1.60435 $\pm$ 0.00001$^*$	\cite{grb071003A_z}	& 07:40:55	\cite{grb071003A_det}	& Oct 05.0771 	\cite{grb071003A_ra846}&	8.46	&	1.7570 $\pm$ 0.0004$^\dagger$	 \\ \hline

%%%new point
071003A &  & & Oct  07.16 \cite{frail2002} &  4.86$^\flat$ &  3.84 $\pm$ 0.01$^\dagger$ \\ \hline

070612A	&	0.617 $\pm$ 0.001$^*$	\cite{grb070612A_z}	& 02:38:41 \cite{grb070612A_det}	& Jun 15.59  	\cite{grb070612A_ra49}&	4.9$^\sharp$	&	3.48 $\pm$ 0.25 \\ \hline

%%%new point
070612A &  &  &   Jun 16.00 \cite{frail2002}  & 8.46 & 3.89 $\pm$ 0.01$^\dagger$ \\ \hline

%%%new point
070125A	&	1.547 $\pm$ 0.001$^{* \circ}$	\cite{grb070125A_z}	& 07:20:42 \cite{grb070125A_det}	&Jan 29.32 \cite{frail2002} & 8.46 & 4.01 $\pm$ 0.01$^\dagger$ \\ \hline

%%%drop this old point
%070125A  &  &   &  Jan 30.2354	\cite{grb070125A_ra}&	8.46	&	4.9294  $\pm$ 0.0004$^\dagger$ \\ \hline

%%%new point
070125A  &  & & Jan 30.95  \cite{grb070125A_ra49} & 4.9$^\sharp$ & 5.64 $\pm$ 0.25 \\ \hline

061121A	&	1.314 $\pm$ 0.001$^*$	\cite{grb061121A_z}	& 15:22:29 \cite{grb061121A_det}	&Nov 22.3825 	\cite{grb061121A_ra}&	8.46	&	0.7419  $\pm$ 0.0004$^\dagger$ \\ \hline

060418A &  1.4901 $\pm$ 0.0001 \cite{grb060418A_z}  & 03:06:08 \cite{grb060418A_det} &  Apr 22.43 \cite{grb060418A_ra} &  8.46$^\flat$ &   4.30 $\pm$ 0.01$^\dagger$ \\ \hline

060218A	&	0.0331 $\pm$ 0.0001$^*$ \cite{grb060218A_z} & 03:34:30 \cite{grb060218A_det}	&  Feb 20.02  \cite{grb060218A_ra}	&	8.46	&	1.87 $\pm$ 0.01$^\dagger$ \\ \hline

%%%new point
060218A & & & Feb 21.97 \cite{grb060218A_ra} & 4.86 & 3.82 $\pm$ 0.01$^\dagger$  \\   \hline

060116A &  6.6 $\pm$ 0.15 \cite{grb060116A_z} & 08:37:27  \cite{grb060116A_det} &   Jan 21.21 \cite{grb060116A_ra} &  8.46$^\sharp$  & 4.85 $\pm$ 0.01$^\dagger$ \\ \hline

051221A	&	0.5465 $\pm$ 0.0001$^*$ \cite{grb051221A_z}	& 01:51:12.976	\cite{grb051221A_det}	& Dec 21.99 	\cite{grb051221A_ra}&	8.46	& 0.91 $\pm$ 0.01$^\dagger$  \\ \hline

051111A	&	1.55 $\pm$ 0.01$^*$ \cite{grb051111A_z}	& 05:59:39 \cite{grb051111A_det}	& Nov 13.15 	\cite{grb051111A_ra}&	8.5$^\sharp$	&	1.90 $\pm$ 0.01$^\dagger$	 \\ \hline

051109A	&	2.346 $\pm$ 0.001$^*$ \cite{grb051109A_z}	&  01:12:20 \cite{grb051109A_det}	&  Nov 11.15 	\cite{grb051109A_ra}&	8.5$^\flat$	&	2.10 $\pm$ 0.01$^\dagger$	 \\ \hline

051022A	&	0.8 $\pm$ 0.1$^*$	\cite{grb051022A_z}$^,$\cite{grb051022A_ra49}	& 13:07:58 \cite{grb051022A_det}& Oct 23.7	\cite{grb051022A_ra49} &	4.9	&	1.2 $\pm$ 0.1	 \\ \hline

%%%new point
051022A  &  & &  Oct 24.09 \cite{grb051022A_ra85}  & 8.5 &  1.54 $\pm$ 0.01$^\dagger$ \\ \hline

050922C &  2.199 $\pm$ 0.001$^{* \circ}$ \cite{grb050922C_z}  &  19:55:54.480 \cite{grb050922C_det} &Sep 24.17 \cite{grb050922C_ra}  & 8.46$^\sharp$ & 1.34 $\pm$ 0.01$^\dagger$ \\ \hline

%050904A	&	6.295 $\pm$ 0.01 \cite{grb050904A_z}&  01:51:44 \cite{grb050904A_det} & Oct 08.26	\cite{grb050904A_ra}&	8.46	&	34.10 $\pm$ 0.01$^\dagger$	 \\ \hline

050904A	&	6.29 $\pm$ 0.01 \cite{grb050904A_z}&  01:51:44 \cite{grb050904A_det} & Oct 09.34	\cite{grb050904A_ra}&	8.46	&	35.26 $\pm$ 0.01$^\dagger$	 \\ \hline

050820A	&	2.6147 $\pm$ 0.0001$^*$ \cite{grb050820A_z}	&   06:34:53 \cite{grb050820A_det}	& Aug 21.20	\cite{grb050820A_ra}&	8.46	&	0.93 $\pm$ 0.01$^\dagger$\\ \hline

%%%new point
050820A  &  & & Aug 22.42 \cite{frail2002} & 4.86$^\sharp$ &  2.15 $\pm$ 0.01$^\dagger$ \\ \hline

\end{tabular}
\end{center}
\caption{Gamma ray burst and radio afterglow data from mid-2005 through March, 2009.}
\end{table}

\pagebreak

\begin{table}[hp]
\begin{center}
\begin{tabular}{|l|l|l|l|l|l|}
\hline
GRB	&	Redshift		&	GRB Time		&	RA Time    &   Frequency  &  Time Lag \\ 
	&				&		(UT)		& 	(UT)		&  (GHz)     &  (d) \\ \hline

050730A	&	3.97 $\pm$ 0.01$^*$ \cite{grb050730A_z}	&  19:58:23 \cite{grb050730A_det}& Aug 02.03	\cite{grb050730A_ra}&	8.5	&2.20 $\pm$ 0.01$^\dagger$ \\ \hline

050724A	&	0.258 $\pm$ 0.002	\cite{grb050724A_z}	& 12:34:09 \cite{grb050724A_det} &	 Jul 25.09	\cite{grb050724A_ra}&	8.46	&0.57 $\pm$ 0.01$^\dagger$	 \\ \hline

050603A	&	2.821 $\pm$ 0.001$^*$	\cite{grb050603A_z}	& 06:29:00.767	\cite{grb050603A_det}	& Jun 03.62	\cite{grb050603A_ra}&	8.5	& 0.35 $\pm$ 0.01$^\dagger$ \\ \hline

%%%new point
050603A & &  &  Jun 09.73 \cite{frail2002} &  4.86$^\flat$&  6.46 $\pm$ 0.01$^\dagger$    \\ \hline

050525A	&	0.606 $\pm$ 0.001$^*$	\cite{grb050525A_z}	&  00:02:53 \cite{grb050525A_det}	&May 25.42	\cite{grb050525A_ra}&	22.5	&	0.42 $\pm$ 0.01$^\dagger$	 \\ \hline

%%%new point
050525A	& &  &  May  27.10 \cite{frail2002} & 15$^\sharp$ & 2.10 $\pm$ 0.01$^\dagger$  \\ \hline

%%%new point
050525A	& &  & May  28.33 \cite{frail2002} & 8.46$^\sharp$ &   3.33 $\pm$ 0.01$^\dagger$  \\ \hline

050416A	&	0.6535 $\pm$ 0.0002 \cite{grb050416A_z} & 11:04:44.5	\cite{grb050416A_det}	& Apr 22.04 \cite{grb050416A_ra}&4.86$^\flat$ &5.58 $\pm$ 0.01$^\dagger$	 \\ \hline

%%%new point
050416A	&  &  &  Apr 28.28 \cite{frail2002} & 8.46$^\flat$&  11.82 $\pm$ 0.01$^\dagger$  \\ \hline

050401A	&	2.9 $\pm$ 0.01$^*$	\cite{grb050401A_z}	& 14:20:11 \cite{grb050401A_det}	& Apr 07.29	\cite{grb050401A_ra}	&	8.5$^\sharp$	&	5.69 $\pm$ 0.01$^\dagger$	 \\ \hline

050315A	&	1.949 $\pm$ 0.001$^*$	\cite{grb050315A_z}	&20:59:42	\cite{grb050315A_det}	&Mar 16.68	\cite{grb050315A_ra}	&	8.5$^\flat$	&	0.81 $\pm$ 0.01$^\dagger$  \\ \hline

031203A	&	0.105 $\pm$ 0.001$^*$	\cite{grb031203A_z}$^,$\cite{grb031203A_ra} &  22:01:28 \cite{grb031203A_det}&  Dec 05.52 	\cite{grb031203A_ra}	&	8.46	& 1.60 $\pm$ 0.01$^\dagger$ \\ \hline

%%%new point
031203A  & & &  Dec 08.35 \cite{grb031203A_ra} & 4.86 & 4.43 $\pm$ 0.01$^\dagger$  \\ \hline

%%%new point
031203A  & & &  Dec 17.38 \cite{grb031203A_ra} & 22.5 &  13.46 $\pm$ 0.01$^\dagger$  \\ \hline

%%%new point
031203A  & & &  Jan 12.29   \cite{grb031203A_ra} & 1.43 & 39.37  $\pm$ 0.01$^\dagger$  \\ \hline

030329A	&	0.1687 $\pm$ 0.0001$^*$	\cite{grb030329A_z}	&  11:36:58 \cite{grb030329A_det}&	 
Mar 30.06 \cite{grb030329A_ra}	&	8.46	&	0.58 $\pm$ 0.01$^\dagger$	 \\ \hline

%%%new point
030329A	&  &  &  Mar 30.53 \cite{grb030329A_ra}  &  4.86$^\flat$ & 1.05 $\pm$ 0.01$^\dagger$ \\ \hline

%%%new point
030329A	& & & Apr 01.13 \cite{grb030329A_ra}  & 15 & 2.65 $\pm$ 0.01$^\dagger$    \\ \hline

%%%new point
030329A	& & & Apr 01.13 \cite{grb030329A_ra}  & 22.5 & 2.65 $\pm$ 0.01$^\dagger$ \\ \hline

%%%new point
030329A	& & & Apr 01.13 \cite{grb030329A_ra}  & 43.3 & 2.65 $\pm$ 0.01$^\dagger$ \\ \hline

%%%new point
030329A	& & & Jun 04.01  \cite{grb030329A_ra}  & 1.43 & 66.53 $\pm$ 0.01$^\dagger$ \\ \hline

%030227A & 1.39 $\pm$ 0.06  \cite{grb030227A_z}  &   08:42:00  \cite{grb030227A_det}  &  Mar  05.97 \cite{grb030227A_ra}  &8.46 &  6.61$\pm$ 0.01$^\dagger$ \\ \hline

030226A & 1.986 $\pm$ 0.001$^\circ$ \cite{grb030226A_z} &   03:46:31.99  \cite{grb030226A_det}  &  Feb 27.25 \cite{grb030226A_ra}  & 8.46  & 1.09 $\pm$ 0.01$^\dagger$ \\ \hline

%%%new point
030226A & & & Mar 10.41 \cite{grb030226A_ra}  & 22.5$^\flat$ & 12.25 $\pm$ 0.01$^\dagger$ \\ \hline

021004A	&	2.331 $\pm$ 0.004	\cite{grb021004A_z1}$^,$\cite{grb021004A_z2} 	&	  12:06:13.57	\cite{grb021004A_det}	& Oct 05.29	\cite{grb021004A_ra}&	22.5	&	0.79 $\pm$ 0.01$^\dagger$ \\ \hline

%%%new point
021004A &  & & Oct 05.29 \cite{frail2002} &  8.46$^\flat$ & 0.79 $\pm$ 0.01$^\dagger$  \\ \hline
 
%%%new point
021004A & & & Oct 08.04 \cite{grb021004A_ra15}  &  15$^\sharp$ & 3.54 $\pm$ 0.08 \\ \hline

%%%new point
021004A & & & Oct 10.17 \cite{grb021004A_ra486} & 4.86 & 5.67 $\pm$ 0.01$^\dagger$  \\ \hline

%XRF 020903A	&	0.25 $\pm$ 0.01	\cite{grb020903A_z}	&	 10:05:37	\cite{grb020903A_det}	&Sep 27.22  \cite{grb020903A_ra}	&	8.46	& 23.80 $\pm$ 0.01$^\dagger$ \\ \hline

020813A	&	1.2545 $\pm$ 0.0001$^*$ \cite{grb020813A_z}	&   02:44:19.17 \cite{grb020813A_det}	&Aug 14.36 \cite{grb020813A_ra}$^,$\cite{grb020813A_ra2} & 8.46   & 1.25 $\pm$ 0.01$^\dagger$	 \\ \hline

%%%new point
020813A &  & & Aug 15.25 \cite{grb020813A_ra2} & 4.86 & 2.14 $\pm$ 0.01$^\dagger$  \\ \hline

\end{tabular}
\end{center}
\caption{Gamma ray burst and radio afterglow data from mid-2002 through mid-2005.}
\end{table}
\bigskip

\pagebreak

\begin{table}[hp]
\begin{center}
\begin{tabular}{|l|l|l|l|l|l|}
\hline
GRB	&	Redshift		&	GRB Time		&	RA Time	&	Frequency  &	Time Lag \\ 
	&				&		(UT)		& 	(UT)		&  (GHz)     &  (d) \\ \hline
020405A	&	0.695 $\pm$ 0.005	\cite{grb020405A_z}	&	00:41:26	\cite{grb020405A_det}	&Apr 06.22	\cite{grb020405A_ra}&	8.46	&	1.19 $\pm$ 0.01$^\dagger$ \\ \hline

%%%new point
020405A	& & & Apr 08.38 \cite{frail2002} & 4.86$^\sharp$ & 3.35 $\pm$ 0.01$^\dagger$ \\ \hline

011211A & 2.142 $\pm$ 0.001$^*$ \cite{grb011211A_z}  & 19:09:21 \cite{grb011211A_det} & Dec   18.58   \cite{grb011211A_ra} & 8.46 & 6.78 $\pm$ 0.01$^\dagger$  \\ \hline

%%%new point
011211A & & & Dec 19.56 \cite{grb011211A_ra} & 22.5$^\sharp$ & 7.76  $\pm$ 0.01$^\dagger$  \\ \hline

%011130A	&	0.5 $\pm$ 0.01$^*$	\cite{grb011130A_z}	&  06:19:35	\cite{grb011130A_det}	&Dec 03.73	\cite{grb011130A_ra}	&	4.86	&	3.47 $\pm$ 0.01$^\dagger$	 \\ \hline

011121A	&	0.362 $\pm$ 0.001	\cite{grb011121A_z}	& 18:47:21	\cite{grb011121A_det}	& Nov 22.83	\cite{grb011121A_ra} & 8.70	& 1.05 $\pm$ 0.01$^\dagger$  \\ \hline

011121A &  & & Nov 25.20 \cite{grb011121A_ra} & 4.80$^\sharp$ & 3.42 $\pm$ 0.01$^\dagger$ \\ \hline

010921A &  0.450 $\pm$ 0.005  \cite{grb010921A_z} &  05:15:50.56   \cite{grb010921A_det} & Oct  17.15  \cite{grb010921A_ra}  & 4.86  & 25.93 $\pm$ 0.01$^\dagger$  \\ \hline

010921A & & & Oct 17.15    \cite{grb010921A_ra}  & 8.46 & 25.93 $\pm$ 0.01$^\dagger$  \\  \hline

010921A &  & & Oct 17.15    \cite{grb010921A_ra}  & 22.5$^\sharp$ & 25.93 $\pm$ 0.01$^\dagger$  \\  \hline

010222A	&	1.477 $\pm$ 0.001$^*$	\cite{grb010222A_z}	&  07:23:30	\cite{grb010222A_det}	&Feb 22.62 \cite{grb010222A_ra}&	22$^\flat$	&	0.31 $\pm$ 0.01$^\dagger$	 \\ \hline

010222A  & & & Feb 23.66 \cite{grb010222A_ra2} & 8.46 &  1.35 $\pm$ 0.01$^\dagger$  \\ \hline

010222A  & & & Feb 24.56 \cite{grb010222A_ra2} & 4.86$^\sharp$ & 2.25  $\pm$ 0.01$^\dagger$  \\ \hline

000926A	&	2.066 $\pm$ 0.001$^{* \circ}$	\cite{grb000926A_z}	&  23:49:33	\cite{grb000926A_det}	&Sep 28.17 \cite{grb000926A_ra}	&	8.46	&	1.18 $\pm$ 0.01$^\dagger$	 \\ \hline

000926A & & & Sep 29.726 \cite{grb000926A_ra} & 4.86 & 2.733 $\pm$ 0.001$^\dagger$  \\ \hline

000926A & & & Oct 04.186  \cite{grb000926A_ra} & 22.5 & 7.193 $\pm$ 0.001$^\dagger$ \\ \hline

000911A	&	1.0585 $\pm$ 0.0001	\cite{grb000911A_z}	&   07:15:25	\cite{grb000911A_det}	&Sep 14.36 	\cite{grb000911A_z}	&	8.46	&	3.06 $\pm$ 0.01$^\dagger$	 \\ \hline

%%%new point
000911A & & &  Sep 22.35 \cite{grb000911A_z} & 4.86$^\sharp$ & 11.05 $\pm$ 0.01$^\dagger$  \\ \hline

000418A	&	1.1181 $\pm$ 0.0001	\cite{grb000418A_z}	&   09:53:10	\cite{grb000418A_det}	& Apr 29.07	\cite{grb000418A_ra}	&	8.46	&	10.66 $\pm$ 0.01$^\dagger$	 \\ \hline

%%%new point
000418A	& & &  May 03.04  \cite{grb000418A_ra}  & 4.86  &  14.63 $\pm$ 0.01$^\dagger$  \\ \hline

%%%new point
000418A	&  & & May 03.04 \cite{grb000418A_ra} &  22.46 &  14.63 $\pm$ 0.01$^\dagger$ \\ \hline

%%%new point
000418A	&  & & May 07.18 \cite{grb000418A_ra} &  8.35  &  18.77  $\pm$ 0.01$^\dagger$ \\ \hline

%%%new point
000301C	&	2.0335 $\pm$ 0.0003	\cite{grb000301C_z}	&  09:51:37	\cite{grb000301C_det}&  Mar 04.98 \cite{grb000301C_ra2} & 15.0$^\flat$ & 3.57 $\pm$ 0.01$^\dagger$  \\ \hline

%%%new point
000301C  &  & &  Mar 05.67   \cite{grb000301C_ra2} & 4.86$^\flat$ & 4.26	 $\pm$ 0.01$^\dagger$  \\ \hline

%%%new point
000301C  & & &  Mar 05.67   \cite{grb000301C_ra2} & 8.46 & 4.26	 $\pm$ 0.01$^\dagger$   \\ \hline

%%%new point
000301C  &  & &  Mar 05.67   \cite{grb000301C_ra2} & 22.5$^\flat$ & 4.26	 $\pm$ 0.01$^\dagger$ \\  \hline

991216A	&	1.00 $\pm$ 0.02	\cite{grb991216A_z}	&  16:07:01	\cite{grb991216A_det}	&Dec  17.783	\cite{grb991216A_ra}	&	4.8	&	1.11 $\pm$ 0.13	 \\ \hline

%%%new point
991216A	&  &  & Dec 18.00  \cite{grb991216A_ra2} &  15$^\flat$ & 1.33 $\pm$ 0.01$^\dagger$ \\ \hline

%%%new point
991216A	&  & & Dec 18.16   \cite{grb991216A_ra2} & 8.46 & 1.49 $\pm$ 0.01$^\dagger$  \\ \hline

%%%new point
991216A	& & & Dec 18.32   \cite{grb991216A_ra2} & 8.42 & 1.65 $\pm$ 0.01$^\dagger$ \\ \hline

%%%new point
991216A	&  & &  Jan 03.11 \cite{grb991216A_ra2} &4.86$^\flat$ & 17.44 $\pm$ 0.01$^\dagger$ \\ \hline

\end{tabular}
\end{center}
\caption{Gamma ray burst and radio afterglow data from mid-1999 through mid-2002.}
\end{table}
\bigskip

\pagebreak

\begin{table}[h]
\begin{center}
\begin{tabular}{|l|l|l|l|l|l|}
\hline
GRB	&	Redshift		&	GRB Time		&	RA Time	&	Frequency  &	Time Lag \\ 
	&				&		(UT)		& 	(UT)		&  (GHz)     &  (d) \\ \hline

991208A	&	0.707 $\pm$ 0.002	\cite{grb991208A_z}	&   04:36:52	\cite{grb991208A_det}	& Dec 10.92  \cite{grb991208A_ra}   & 4.86 & 2.73 $\pm$ 0.01$^\dagger$  \\ \hline

%%%new point
991208A & & & Dec 10.92  \cite{grb991208A_ra}   & 8.46 & 2.73 $\pm$ 0.01$^\dagger$ \\ \hline

%%%new point
991208A & & & Dec 11.51 \cite{grb991208A_ra}	&	15.0	&	3.32 $\pm$ 0.01$^\dagger$	 \\ \hline

%%%new point
991208A & & &  Dec 14.84  \cite{grb991208A_ra}	& 30$^\sharp$ &  6.65 $\pm$ 0.01$^\dagger$	 \\ \hline

%%%new point
991208A & & &  Dec 15.90  \cite{grb991208A_ra} &  1.43$^\flat$ & 7.71 $\pm$ 0.01$^\dagger$ \\ \hline

%%%new point
991208A & & & Dec 21.96  \cite{grb991208A_ra}	&	14.97 	&  13.77 $\pm$ 0.01$^\dagger$	 \\ \hline

%%%new point
991208A & & & Dec 21.96  \cite{grb991208A_ra}	&	22.49 	&  13.77 $\pm$ 0.01$^\dagger$	 \\ \hline

990510A	&	1.619 $\pm$ 0.002$^\circ$	\cite{grb990510A_ra}	&  08:49	\cite{grb990510A_det}	&  May 13.68	\cite{grb990510A_ra}	&	8.7	&	3.31 $\pm$ 0.19	 \\ \hline

%%%new point
990510A  & & & May 19.59 \cite{grb990510A_ra} &  4.8$^\flat$ & 9.22 $\pm$ 0.24 \\ \hline

%%%new point
990510A  & & & May 19.59 \cite{grb990510A_ra} &  8.6$^\flat$ & 9.22 $\pm$ 0.24 \\ \hline

990506A	&	1.30658 $\pm$ 0.00004	\cite{grb000418A_z}	&  11:23:31.0	\cite{grb990506A_det}	& May  08.13 \cite{grb990506A_ra}	&	8.46	&	1.66 $\pm$ 0.01$^\dagger$	 \\ \hline

990123A	&	1.61 $\pm$ 0.01$^*$	\cite{grb990123A_z}	& 09:46:56.1	\cite{grb990123A_det}	& Jan 24.65	\cite{grb990123A_ra1}$^,$\cite{grb990123A_ra2}	&	8.46	&	1.24 $\pm$ 0.01$^\dagger$	 \\ \hline

980703A	&	0.967 $\pm$ 0.001$^*$	\cite{grb980703A_z}	& 04:22:45	\cite{grb980703A_det}	&   Jul 04.40	\cite{grb980703A_ra}	&	4.86	&	1.22 $\pm$ 0.01$^\dagger$	 \\ \hline

%%%new point
980703A & & & Jul 07.35 \cite{grb980703A_ra}  & 8.46  & 4.17 $\pm$ 0.01$^\dagger$ \\ \hline

%%%new point
980703A & & &  Jul 08.49 \cite{grb980703A_ra}  & 1.43$^\sharp$ & 5.31 $\pm$ 0.01$^\dagger$  \\ \hline

%980425A	&	0.0085 $\pm$ 0.0002	\cite{grb980425A_z}	& 21:49:09	\cite{grb980425A_det}	&Apr 28.73	\cite{grb980425A_ra}	&	4.996540967	&	2.82 $\pm$ 0.01$^\dagger$	 \\ \hline

%970828A	&	0.9578 $\pm$ 0.0001	\cite{grb970828A_z}	& 17:44:36	\cite{grb970828A_det}	&Aug 31.19 \cite{grb970828A_ra}	&	4.996540967	&	2.45$\pm$ 0.01 $^\dagger$	 \\ \hline

970828A	&	0.9578 $\pm$ 0.0001	\cite{grb970828A_z}	& 17:44:36	\cite{grb970828A_det}	& Sep 01.27  \cite{grb970828A_z}	&	8.46$^\flat$	&	3.53 $\pm$ 0.01$^\dagger$	 \\ \hline

970508A	&	0.835 $\pm$ 0.001$^{* \circ}$	\cite{grb970508A_z1}$^,$\cite{grb970508A_z2}$^,$\cite{grb970508A_z3}	&  21:42	\cite{grb970508A_det}	& May 13.96	\cite{grb970508A_ra}	&	8.46	&	5.06 $\pm$ 0.01$^\dagger$ \\ \hline

%%%new point
970508A	&  & & May   15.09 \cite{grb970508A_ra}	& 1.43$^\sharp$ & 6.19 $\pm$ 0.01$^\dagger$ \\ \hline

%%%new point
970508A	& & & May 15.13 \cite{grb970508A_ra}	&  4.86 & 6.23 $\pm$ 0.01$^\dagger$ \\ \hline

\end{tabular}
\end{center}
\caption{Gamma ray burst and radio afterglow data from 1997 through mid-1999.}
\end{table}


\begin{thebibliography}{1}

\bibitem{concord1}
D.~N.~Spergel {\it et al.}  [WMAP Collaboration],
 %  ``First Year Wilkinson Microwave Anisotropy Probe (WMAP) Observations:
  %Determination of Cosmological Parameters,''
  Astrophys.\ J.\ Suppl.\  {\bf 148}, 175 (2003). 
  %%CITATION = ASTRO-PH 0302209;%%

\bibitem{concord2}
M.~Tegmark {\it et al.}  [SDSS Collaboration],
  %``Cosmological parameters from SDSS and WMAP,''
  Phys.\ Rev.\ D {\bf 69}, 103501 (2004). 
  %%CITATION = ASTRO-PH 0310723;%%

\bibitem{Komatsu:2008hk}
  E.~Komatsu {\it et al.}  [WMAP Collaboration],
  %``Five-Year Wilkinson Microwave Anisotropy Probe (WMAP\altaffilmark 1 )
  %Observations:Cosmological Interpretation,''
  Astrophys.\ J.\ Suppl.\  {\bf 180}, 330 (2009). 
%  [arXiv:0803.0547 [astro-ph]].
  %%CITATION = APJSA,180,330;%%

\bibitem{svg}
  S.~Gardner,
  %``Observing Dark Matter via the Gyromagnetic Faraday Effect,''
  Phys.\ Rev.\ Lett.\  {\bf 100}, 041303 (2008). 
%  [arXiv:astro-ph/0611684].
  %%CITATION = PRLTA,100,041303;%%

\bibitem{Ullio:2002pj}
  P.~Ullio, L.~Bergstrom, J.~Edsjo, and C.~G.~Lacey,
  %``Cosmological dark matter annihilations into gamma-rays: A closer look,''
  Phys.\ Rev.\  D {\bf 66}, 123502 (2002). 
%  [arXiv:astro-ph/0207125].
  %%CITATION = PHRVA,D66,123502;%%

\bibitem{hall} 
J.~L. Hall, in {\em Atomic Masses and Fundamental Constants}, 
edited by J.~H. Sanders and A.~H. Wapstra (Plenum, New York, 1976), p. 322. 
%R.~H. Garstang, Astrophys. J. 79, 1260 (1974).

\bibitem{crab} B. Warner and R. Nather, Nature {\bf 222}, 157 (1969). 
%%CITATION = NATUA,222,157;%%

\bibitem{schprl} 
  B.~E.~Schaefer,
  %``Severe Limits on Variations of the Speed of Light with Frequency,''
  Phys.\ Rev.\ Lett.\  {\bf 82}, 4964 (1999).
% and references therein.
%  [arXiv:astro-ph/9810479].
  %%CITATION = PRLTA,82,4964;%%

\bibitem{AmelinoCamelia:1997gz}
  G.~Amelino-Camelia {\em et al.}, 
%, J.~R.~Ellis, N.~E.~Mavromatos, D.~V.~Nanopoulos, and S.~Sarkar,
  %``Potential Sensitivity of Gamma-Ray Burster Observations to Wave Dispersion
  %in Vacuo,''
  Nature {\bf 393}, 763 (1998). 
%  [arXiv:astro-ph/9712103].
  %%CITATION = NATUA,393,763;%%

\bibitem{abdo}
A. A. Abdo {\it et al.} [Fermi LAT Collaboration],
%``A limit on the variation of the speed of light arising from quantum gravity effects,"
 Nature {\bf 462}, 331 (2009).

\bibitem{GGT}
M.~Gell-Mann, M.~L.~Goldberger, and W.~E.~Thirring, Phys.\ Rev.\ {\bf 95}, 1612 (1954). 

\bibitem{GGT2}
M.~L.~Goldberger, Phys.\ Rev.\ {\bf 97}, 508 (1955). 
%R.~E.~Prange, Phys. Rev. {\bf 110}, 240 (1958).

\bibitem{Hemmert:1997tj}
  T.~R.~Hemmert, B.~R.~Holstein, J.~Kambor, and G.~Kn\"ochlein,
  %``Compton scattering and the spin structure of the nucleon at low
  %energies,''
  Phys.\ Rev.\  D {\bf 57}, 5746 (1998). 
%  [arXiv:nucl-th/9709063].
  %%CITATION = PHRVA,D57,5746;%%

\bibitem{let1} W.~E.~Thirring, Philos.\ Mag.\ {\bf 41}, 1193 (1950). 

\bibitem{fermi} E. Fermi, 
{\em Nuclear Physics, Revised Edition} (University of Chicago Press, Chicago, IL, 1974), p. 202.

\bibitem{newton} R.~G.~Newton, {\em Scattering Theory of Waves and Particles,
1st Edition} (McGraw-Hill, New York, 1966), p. 24ff. 

\bibitem{ps} M.~E.~Peskin and D.~V.~Schroeder, {\em An Introduction to Quantum Field Theory}
(Addison-Wesley Publishing Company, Reading, MA, 1995). 

\bibitem{sikivieyang} 
  P.~Sikivie and Q.~Yang,
  %``Bose-Einstein Condensation of Dark Matter Axions,''
  Phys.\ Rev.\ Lett.\  {\bf 103}, 111301 (2009). 
%  [arXiv:0901.1106 [hep-ph]].
  %%CITATION = PRLTA,103,111301;%%

\bibitem{lapidus} L.~I.~Lapidus and Chou Kuang-Chao, J.\ Exptl.\ Theoret.\ Phys.\ (USSR) 
{\bf 39}, 1286 (1960) 
[Sov.\ Phys.\ JETP {\bf 12}, 898 (1961)]. 

\bibitem{brodsky}
S.~J.~Brodsky and J.~R.~Primack,
%The Electromagnetic Interactions of Composite Systems.  
Ann.\ Phys.\  {\bf 52}, 315 (1969).

\bibitem{ellis} 
  J.~R.~Ellis {\em et al.}, 
%K.~Farakos, N.~E.~Mavromatos, V.~A.~Mitsou, and D.~V.~Nanopoulos,
  %``Astrophysical probes of the constancy of the velocity of light,''
  Astrophys.\ J.\  {\bf 535}, 139 (2000). 
%  [arXiv:astro-ph/9907340].
  %%CITATION = ASJOA,535,139;%%

\bibitem{jacobpiran} 
  U.~Jacob and T.~Piran,
%  %``Lorentz-violation-induced arrival delays of cosmological particles,''
  JCAP {\bf 0801}, 031 (2008). 
%%  [arXiv:0712.2170 [astro-ph]].
%  %%CITATION = JCAPA,0801,031;%%

\bibitem{mirror}
  B.~Holdom,
  %``Two U(1)'S And Epsilon Charge Shifts,''
  Phys.\ Lett.\  B {\bf 166}, 196 (1986).
  %%CITATION = PHLTA,B166,196;%%

\bibitem{feldman}
 D.~Feldman, Z.~Liu, and P.~Nath,
  %``The Stueckelberg Z' extension with kinetic mixing and milli-charged dark
  %matter from the hidden sector,''
  Phys.\ Rev.\  D {\bf 75}, 115001 (2007).
  
\bibitem{kusenko}
A.~Kusenko and P.~J.~Steinhardt,
  %``Q-ball candidates for self-interacting dark matter,''
  Phys.\ Rev.\ Lett.\  {\bf 87}, 141301 (2001).

\bibitem{supp} See appended supplementary
information.

\bibitem{shoemaker}
I.~M.~Shoemaker and A.~Kusenko,
  %``The ground states of baryoleptonic Q-balls in supersymmetric models,''
  Phys.\ Rev.\  D {\bf 78}, 075014 (2008).

\bibitem{davidson}
  S.~Davidson, S.~Hannestad, and G.~Raffelt,
  %``Updated bounds on milli-charged particles,''
%J.\ High Energy Phys.\ 
 JHEP {\bf 0005}, 003 (2000). 
%  [arXiv:hep-ph/0001179].
  %%CITATION = JHEPA,0005,003;%%

\bibitem{Ahlers:2007qf}
  M.~Ahlers {\em et al.}, 
%, H.~Gies, J.~Jaeckel, J.~Redondo, and A.~Ringwald,
  %``Laser experiments explore the hidden sector,''
  Phys.\ Rev.\  D {\bf 77}, 095001 (2008). 
%  [arXiv:0711.4991 [hep-ph]].
  %%CITATION = PHRVA,D77,095001;%%

\bibitem{Melchiorri:2007sq}
  A.~Melchiorri, A.~Polosa, and A.~Strumia,
  %``New bounds on millicharged particles from cosmology,'' %CMB
  Phys.\ Lett.\  B {\bf 650}, 416 (2007). 
%  [arXiv:hep-ph/0703144].
  %%CITATION = PHLTA,B650,416;%%

\bibitem{burrage}
C.~Burrage, J.~Jaeckel, J.~Redondo, and A.~Ringwald,
  %``Late time CMB anisotropies constrain mini-charged particles,''
  JCAP {\bf 11}, 002 (2009).

\bibitem{Gradwohl:1992ue}
  B.~A.~Gradwohl and J.~A.~Frieman,
  %``Dark Matter, Long Range Forces, And Large Scale Structure,''
  Astrophys.\ J.\  {\bf 398}, 407 (1992).
  %%CITATION = ASJOA,398,407;%%

\bibitem{acker}
  L.~Ackerman, M.~R.~Buckley, S.~M.~Carroll, and M.~Kamionkowski,
  %``Dark Matter and Dark Radiation,''
  Phys.\ Rev.\  D {\bf 79}, 023519 (2009).
%  [arXiv:0810.5126 [hep-ph]].
  %%CITATION = PHRVA,D79,023519;%%

\bibitem{feng}
J.~L.~Feng, M.~Kaplinghat, H.~Tu, and H.~B.~Yu,
  %``Hidden Charged Dark Matter,''
  JCAP {\bf 07}, 004 (2009).

\bibitem{massoredondo} E.~Mass{\'o} and J.~Redondo, Phys.\ Rev.\ Lett.\ 
{\bf 97}, 151802 (2006). 

\bibitem{burgess}
  C.~P.~Burgess {\em et al.}, 
%J.~P.~Conlon, L.~Y.~Hung, C.~H.~Kom, A.~Maharana and F.~Quevedo,
  %``Continuous Global Symmetries and Hyperweak Interactions in String
  %Compactifications,''
  JHEP {\bf 07}, 073 (2008).

\bibitem{Bombelli:2004tq}
  L.~Bombelli and O.~Winkler,
  %``Comparison of QG-induced dispersion with standard physics effects,''
  Class.\ Quant.\ Grav.\  {\bf 21}, L89 (2004). 
%  [arXiv:gr-qc/0403049].
  %%CITATION = CQGRD,21,L89;%%

\bibitem{llanta} E.~Llanta and R.~Tarrach, 
 Phys.\ Lett.\  B {\bf 78}, 586 (1978). 

\bibitem{leew} B.~W.~Lee and S.~Weinberg, 
  Phys.\ Rev.\ Lett.\  {\bf 39}, 165 (1977). 

\bibitem{campbell}
G.~K.~Campbell {\em et~al.}, 
% Photon recoil momentum in dispersive media. 
Phys.\ Rev.\ Lett.\ {\bf 94}, 170403 (2005). 
%Even with $M=10^{-6}\,\hbox{eV}$, the cosmological dark-matter number density
%is still smaller than the number density of the dilute system studied here. 
\bigskip

\pagebreak

\section{References}
%\begin{thebibliography}{100}
%\setcounter{enumi}{36}

\bibitem{grb090328A_z}
{Cenko}, S.~B. \emph{et~al.}
\newblock {GRB 090328: Gemini South redshift}.
\newblock {\em GCN Circular} 9053, http://gcn.gsfc.nasa.gov/gcn3/9053.gcn3
  (2009).

\bibitem{grb090328A_det}
{McEnery}, J. \emph{et~al.}
\newblock {Fermi LAT and GBM detections of GRB 090328}.
\newblock {\em GCN Circular} 9044, http://gcn.gsfc.nasa.gov/gcn3/9044.gcn3
  (2009).

\bibitem{grb090328A_ra}
{Frail}, D.~A. \emph{et~al.}
\newblock {GRB 090328: Radio afterglow detection}.
\newblock {\em GCN Circular} 9060, http://gcn.gsfc.nasa.gov/gcn3/9060.gcn3
  (2009).

\bibitem{grb090323A_z}
{Chornock}, R. \emph{et~al.}
\newblock {GRB 090323 Gemini-South redshift}.
\newblock {\em GCN Circular} 9028, http://gcn.gsfc.nasa.gov/gcn3/9028.gcn3
  (2009).

\bibitem{grb090323A_det}
{Ohno}, M. \emph{et~al.}
\newblock {Fermi GBM and LAT detections of GRB 090323}.
\newblock {\em GCN Circular} 9021, http://gcn.gsfc.nasa.gov/gcn3/9021.gcn3
  (2009).

\bibitem{grb090323A_ra}
Harrison, F., Cenko, B., Frail, D.~A., Chandra, P. \& Kulkarni, S.
\newblock {GRB 090323: Radio afterglow detection}.
\newblock {\em GCN Circular} 9043, http://gcn.gsfc.nasa.gov/gcn3/9043.gcn3
  (2009).

\bibitem{grb090313A_z}
Chornock, R. \emph{et~al.}
\newblock GRB 090313: Gemini-S redshift.
\newblock {\em GCN Circular} 8994, http://gcn.gsfc.nasa.gov/gcn3/8994.gcn3
  (2009).

\bibitem{grb090313A_det}
Mao, J. \emph{et~al.}
\newblock {GRB 090313: Swift detection of a burst}.
\newblock {\em GCN Circular} 8980, http://gcn.gsfc.nasa.gov/gcn3/8980.gcn3
  (2009).

\bibitem{grb090313A_ra}
{Pooley}, G.  \emph{et~al.}
\newblock {Radio detection of GRB 090313}.
\newblock {\em GCN Circular} 9003, http://gcn.gsfc.nasa.gov/gcn3/9003.gcn3
  (2009).

\bibitem{grb081007A_z}
{Berger}, E. \emph{et~al.}
\newblock {GRB 081007: Gemini-south redshift}.
\newblock {\em GCN Circular} 8335, http://gcn.gsfc.nasa.gov/gcn3/8335.gcn3
  (2008).

\bibitem{grb081007A_det}
{Baumgartner}, W.~H. \emph{et~al.}
\newblock {GRB 081007: Swift detection of a burst with optical afterglow}.
\newblock {\em GCN Circular} 8330, http://gcn.gsfc.nasa.gov/gcn3/8330.gcn3
  (2008).

\bibitem{grb081007A_ra}
{Soderberg}, A.  \emph{et~al.}
\newblock {GRB 081007: Radio detection}.
\newblock {\em GCN Circular} 8354, http://gcn.gsfc.nasa.gov/gcn3/8354.gcn3
  (2008).

\bibitem{grb080810A_z}
{Prochaska}, J.~X. \emph{et~al.}
\newblock {GRB 080810: Keck/HIRES spectroscopy}.
\newblock {\em GCN Circular} 8083, http://gcn.gsfc.nasa.gov/gcn3/8083.gcn3
  (2008).

\bibitem{grb080810A_det}
{Page}, K.~L. \emph{et~al.}
\newblock {GRB 080810: Swift detection of a burst with a bright optical
  afterglow}.
\newblock {\em GCN Circular} 8080, http://gcn.gsfc.nasa.gov/gcn3/8080.gcn3
  (2008).

\bibitem{grb080810A_ra}
{Chandra}, P. \emph{et~al.}
\newblock {Radio detection of GRB 080810 with the VLA}.
\newblock {\em GCN Circular} 8103, http://gcn.gsfc.nasa.gov/gcn3/8103.gcn3
  (2008).

\bibitem{grb080603A_z}
{Perley}, D.~A. \emph{et~al.}
\newblock {GRB 080603: Gemini-North redshift}.
\newblock {\em GCN Circular} 7791, http://gcn.gsfc.nasa.gov/gcn3/7791.gcn3
  (2008).

\bibitem{grb080603A_det}
{Paizis}, A. \emph{et~al.}
\newblock {GRB 080603: A long GRB detected by INTEGRAL}.
\newblock {\em GCN Circular} 7790, http://gcn.gsfc.nasa.gov/gcn3/7790.gcn3
  (2008).

\bibitem{grb080603A_ra}
{Chandra}, P. \emph{et~al.}
\newblock {VLA detection of INTEGRAL burst GRB 080603A}.
\newblock {\em GCN Circular} 7855, http://gcn.gsfc.nasa.gov/gcn3/7855.gcn3
  (2008).

\bibitem{grb080319B_z}
{Vreeswijk}, P.~M. \emph{et~al.}
\newblock {VLT/UVES redshift of GRB 080319B from FeII fine-structure lines}.
\newblock {\em GCN Circular} 7451, http://gcn.gsfc.nasa.gov/gcn3/7451.gcn3
  (2008).

\bibitem{grb080319B_det}
{Beckmann}, V. \emph{et~al.}
\newblock {GRB 080319A/B/C: INTEGRAL SPI-ACS light curves available}.
\newblock {\em GCN Circular} 7450, http://gcn.gsfc.nasa.gov/gcn3/7450.gcn3
  (2008).

\bibitem{grb080319B_ra}
{Soderberg}, A. \emph{et~al.}
\newblock {Radio detection of GRB 080319B}.
\newblock {\em GCN Circular} 7506, http://gcn.gsfc.nasa.gov/gcn3/7506.gcn3
  (2008).

%\bibitem{grb080109A_z}
%{Malesani}, D. \emph{et~al.}
%\newblock {Transient in NGC 2770: spectroscopic evidence for a SN}.
%\newblock {\em GCN Circular} 7169. http://gcn.gsfc.nasa.gov/gcn3/7169.gcn3
 % (2008).

%\bibitem{grb080109A_det}
%{Ofek}, E.~O. \emph{et~al.}
%\newblock {XRF in NGC 2770: Search for X-ray periodicity}.
%\newblock {\em GCN Circular} 7172. http://gcn.gsfc.nasa.gov/gcn3/7172.gcn3
 % (2008).

%\bibitem{grb080109A_ra}
%{Soderberg}, A. \emph{et~al.}
%\newblock {Discovery of Radio Emission from Transient in NGC 2770}.
%\newblock {\em GCN Circular} 7178. http://gcn.gsfc.nasa.gov/gcn3/7178.gcn3
 % (2008).

\bibitem{grb071122A_z}
{Cucchiara}, A. \emph{et~al.}
\newblock {GRB 071122: Gemini absorption redshift}.
\newblock {\em GCN Circular} 7124, http://gcn.gsfc.nasa.gov/gcn3/7124.gcn3
  (2007).

\bibitem{grb071122A_det}
{Stamatikos}, M. \emph{et~al.}
\newblock {GRB 071122: Swift detection of a burst}.
\newblock {\em GCN Circular} 7121, http://gcn.gsfc.nasa.gov/gcn3/7121.gcn3
  (2007).

\bibitem{grb071122A_ra}
{Chandra}, P.  \emph{et~al.}
\newblock {VLA radio detection of GRB 071122}.
\newblock {\em GCN Circular} 7132, http://gcn.gsfc.nasa.gov/gcn3/7132.gcn3
  (2007).

\bibitem{grb071020A_z}
{Jakobsson}, P. \emph{et~al.}
\newblock {GRB 071020: VLT spectroscopy}.
\newblock {\em GCN Circular} 6952, http://gcn.gsfc.nasa.gov/gcn3/6952.gcn3
  (2007).

\bibitem{grb071020A_det}
{Holland}, S.~T. \emph{et~al.}
\newblock {GRB 071020: Swift detection of a bright burst}.
\newblock {\em GCN Circular} 6949, http://gcn.gsfc.nasa.gov/gcn3/6949.gcn3
  (2007).

\bibitem{grb071020A_ra}
{Chandra}, P.  \emph{et~al.}
\newblock {Radio detection of GRB 071020 with the VLA}.
\newblock {\em GCN Circular} 6978, http://gcn.gsfc.nasa.gov/gcn3/6978.gcn3
  (2007).

\bibitem{grb071010B_z}
{Stern}, D. \emph{et~al.}
\newblock {GRB 071010B: Keck/DEIMOS emission-line redshift}.
\newblock {\em GCN Circular} 6928, http://gcn.gsfc.nasa.gov/gcn3/6928.gcn3
  (2007).

\bibitem{grb071010B_det}
{Markwardt}, C.~B. \emph{et~al.}
\newblock {GRB 071010B: Swift detection of a burst}.
\newblock {\em GCN Circular} 6871, http://gcn.gsfc.nasa.gov/gcn3/6871.gcn3
  (2007).

\bibitem{grb071010B_ra}
{Chandra}, P. \emph{et~al.}
\newblock {VLA detection of radio afterglow of GRB 071010B}.
\newblock {\em GCN Circular} 6915, http://gcn.gsfc.nasa.gov/gcn3/6915.gcn3
  (2007).

\bibitem{grb071003A_z}
Perley, D.~A.  \emph{et~al.}
\newblock {GRB 071003: Broadband follow-up observations of a 
very bright gamma-ray burst in a galactic halo}.
 \newblock{\em Astrophys. J.} {\bf 688}, {470-490} (2008).

\bibitem{grb071003A_det}
{Schady}, P. \emph{et~al.}
\newblock {GRB 071003: Swift detection of a bright burst}.
\newblock {\em GCN Circular} 6837, http://gcn.gsfc.nasa.gov/gcn3/6837.gcn3
  (2007).

\bibitem{grb071003A_ra846}
{Chandra}, P. \emph{et~al.}
\newblock {Radio detection of GRB 071003 with the VLA}.
\newblock {\em GCN Circular} 6853, http://gcn.gsfc.nasa.gov/gcn3/6853.gcn3
  (2007).

\bibitem{frail2002}
Frail, D.
\newblock {Radio afterglow data}.
\newblock {http://www.aoc.nrao.edu/$\sim$dfrail/radio\_data\_2002\_new.dat}  (2002).


\bibitem{grb070612A_z}
{Cenko}, S.~B. \emph{et~al.}
\newblock {GRB 070612A: Gemini spectroscopic redshift}.
\newblock {\em GCN Circular} 6556, http://gcn.gsfc.nasa.gov/gcn3/6556.gcn3
  (2007).

\bibitem{grb070612A_det}
{Uehara}, T. \emph{et~al.}
\newblock {GRB 070612A: Suzaku/WAM observation of the prompt emission}.
\newblock {\em GCN Circular} 6533, http://gcn.gsfc.nasa.gov/gcn3/6533.gcn3
  (2007).

\bibitem{grb070612A_ra49}
v.~d.~{Horst}, A.~J. \emph{et~al.}
\newblock {GRB 070612A: Possible WSRT radio detection}.
\newblock {\em GCN Circular} 6549, http://gcn.gsfc.nasa.gov/gcn3/6549.gcn3
  (2007).


\bibitem{grb070125A_z}
{Fox}, D.~B. \emph{et~al.}
\newblock {GRB 070125: Redshift z $\gtrsim$ 1.54 from Gemini afterglow spectrum}.
\newblock {\em GCN Circular} 6071, http://gcn.gsfc.nasa.gov/gcn3/6071.gcn3
  (2007).

\bibitem{grb070125A_det}
{Bellm}, E. \emph{et~al.}
\newblock {RHESSI Spectrum of GRB 070125}.
\newblock {\em GCN Circular} 6025, http://gcn.gsfc.nasa.gov/gcn3/6025.gcn3
  (2007).

%\bibitem{grb070125A_ra}
%{Chandra}, P.  \emph{et~al.}
%\newblock {Radio detection of GRB 070125}.
%\newblock {\em GCN Circular} 6061. http://gcn.gsfc.nasa.gov/gcn3/6061.gcn3
 % (2007).

\bibitem{grb070125A_ra49}
v.~d.~{Horst}, A.~J. \emph{et~al.}
   \newblock{GRB 070125: WSRT radio detection}.
   \newblock {\em GCN Circular} 6063, http://gcn.gsfc.nasa.gov/gcn3/6063.gcn3 (2007).

\bibitem{grb061121A_z}
{Bloom}, J.~S., {Perley}, D.~A. \& {Chen}, H.~W.
\newblock {GRB 061121: Spectroscopic redshift}.
\newblock {\em GCN Circular} 5826, http://gcn.gsfc.nasa.gov/gcn3/5826.gcn3
  (2006).

\bibitem{grb061121A_det}
{Page}, K.~L. \emph{et~al.}
\newblock {GRB 061121: Swift detection of a bright burst with an optical
  counterpart}.
\newblock {\em GCN Circular} 5823, http://gcn.gsfc.nasa.gov/gcn3/5823.gcn3
  (2006).

\bibitem{grb061121A_ra}
{Chandra}, P.  \emph{et~al.}
\newblock {Radio detection of GRB 061121}.
\newblock {\em GCN Circular} 5843, http://gcn.gsfc.nasa.gov/gcn3/5843.gcn3
  (2006).

\bibitem{grb060418A_z}
{Prochaska}, J.~X. \emph{et~al.}
\newblock {GRB 060418: Further analysis of MIKE spectroscopy}.
\newblock {\em GCN Circular} 5002, http://gcn.gsfc.nasa.gov/gcn3/5002.gcn3
  (2006).

\bibitem{grb060418A_det}
{Falcone}, A.~D. \emph{et~al.}
\newblock {GRB 060418: Swift detection of a burst with bright x-ray and optical
  afterglow}.
\newblock {\em GCN Circular} 4966, http://gcn.gsfc.nasa.gov/gcn3/4966.gcn3
  (2006).

\bibitem{grb060418A_ra}
Frail, D.
\newblock {GRB 060418 data}.
\newblock {http://www.aoc.nrao.edu/$\sim$dfrail/grb060418.dat}  (2006).

\bibitem{grb060218A_z}
{Mirabal}, N.  \emph{et~al.}
\newblock {GRB 060218: MDM redshift}.
\newblock {\em GCN Circular} 4792, http://gcn.gsfc.nasa.gov/gcn3/4792.gcn3
  (2006).

\bibitem{grb060218A_det}
{Cusumano}, G. \emph{et~al.}
\newblock {GRB 060218: Swift-BAT detection of a possible burst}.
\newblock {\em GCN Circular} 4775, http://gcn.gsfc.nasa.gov/gcn3/4775.gcn3
  (2006).

\bibitem{grb060218A_ra}
{Soderberg}, A.~M. \emph{et~al.}
\newblock {Relativistic ejecta from X-ray flash XRF 060218 and the rate of
  cosmic explosions}.
\newblock {\em Nature}{ \bf 442}, 1014--1017  (2006).

\bibitem{grb060116A_z}
{Grazian}, A. \emph{et~al.}
\newblock {GRB 060116: photometric redshift - the farthest GRB?}
\newblock {\em GCN Circular} 4545, http://gcn.gsfc.nasa.gov/gcn3/4545.gcn3
  (2006).

\bibitem{grb060116A_det}
{Campana}, S. \emph{et~al.}
\newblock {GRB 060116: Swift-BAT detection of a burst}.
\newblock {\em GCN Circular} 4519, http://gcn.gsfc.nasa.gov/gcn3/4519.gcn3
  (2006).

\bibitem{grb060116A_ra}
Frail, D.
\newblock {GRB 060116 data}.
\newblock {http://www.aoc.nrao.edu/$\sim$dfrail/grb060116.dat},  (2006).

\bibitem{grb051221A_z}
{Berger}, E. \& {Soderberg}, A.~M.
\newblock {GRB 051221: Redshift from Gemini}.
\newblock {\em GCN Circular} 4384, http://gcn.gsfc.nasa.gov/gcn3/4384.gcn3
  (2005).

\bibitem{grb051221A_det}
{Golenetskii}, S. \emph{et~al.}
\newblock {Konus-Wind observation of GRB 051221A}.
\newblock {\em GCN Circular} 4394, http://gcn.gsfc.nasa.gov/gcn3/4394.gcn3
  (2005).

%\bibitem{grb051221A_ra}
%{Frail}, D.~A. \emph{et~al.}
%\newblock {GRB 051221A: VLA Radio Observations}.
%\newblock {\em GCN Circular} 4416. http://gcn.gsfc.nasa.gov/gcn3/4416.gcn3
 % (2005).

\bibitem{grb051221A_ra}
{Soderberg}, A.~M. {\em et al.}
  \newblock{The afterglow, energetics, and host galaxy of the short-hard gamma-ray burst 051221a}.
 \newblock{ \em Astrophys. J.} {\bf 650}, {261-271} (2006).

\bibitem{grb051111A_z}
{Hill}, G., {Prochaska}, J.~X., {Fox}, D., {Schaefer}, B. \& {Reed}, M.
\newblock {GRB 051111: Keck HIRES redshift}.
\newblock {\em GCN Circular} 4255, http://gcn.gsfc.nasa.gov/gcn3/4255.gcn3
  (2005).

\bibitem{grb051111A_det}
{Yamaoka}, K. \emph{et~al.}
\newblock {GRB 051111: Suzaku WAM observation of the prompt emission}.
\newblock {\em GCN Circular} 4299, http://gcn.gsfc.nasa.gov/gcn3/4299.gcn3
  (2005).

\bibitem{grb051111A_ra}
{Frail}, D.~A. \emph{et~al.}
\newblock {GRB 051111: Radio afterglow}.
\newblock {\em GCN Circular} 4270, http://gcn.gsfc.nasa.gov/gcn3/4270.gcn3
  (2005).

\bibitem{grb051109A_z}
{Quimby}, R., {Fox}, D., {Hoeflich}, P., {Roman}, B. \& {Wheeler}, J.~C.
\newblock {GRB 051109: HET optical spectrum and absorption redshift}.
\newblock {\em GCN Circular} 4221, http://gcn.gsfc.nasa.gov/gcn3/4221.gcn3
  (2005).

\bibitem{grb051109A_det}
{Tagliaferri}, G. \emph{et~al.}
\newblock {GRB 051109: Swift detection of a burst}.
\newblock {\em GCN Circular} 4213, http://gcn.gsfc.nasa.gov/gcn3/4213.gcn3
  (2005).

\bibitem{grb051109A_ra}
{Frail}, D.~A. \emph{et~al.}
\newblock {GRB 051109A}.
\newblock {\em GCN Circular} 4244, http://gcn.gsfc.nasa.gov/gcn3/4244.gcn3
  (2005).

\bibitem{grb051022A_z}
{Gal-Yam}, A. \emph{et~al.}
\newblock {GRB 051022: host galaxy redshift}.
\newblock {\em GCN Circular} 4156, http://gcn.gsfc.nasa.gov/gcn3/4156.gcn3
  (2005).

\bibitem{grb051022A_ra49}
 {Rol}, E. \emph{et~al.}
\newblock{GRB 051022: Physical parameters and extinction of a prototype dark burst}.
\newblock {\em Astrophys. J.}{ \bf 669}, 1098-1106 (2007).

\bibitem{grb051022A_det}
{Olive}, J. \emph{et~al.}
\newblock {GRB 051022 (HETE 3590), A GRB detected by HETE}.
\newblock {\em GCN Circular} 4131, http://gcn.gsfc.nasa.gov/gcn3/4131.gcn3
  (2005).

%\bibitem{grb051022A_ra49}
%{Horst}, A.~J.~v.~d. \emph{et~al.}
%\newblock {GRB 051022: WSRT Radio Detection}.
%\newblock {\em GCN Circular} 4158. http://gcn.gsfc.nasa.gov/gcn3/4158.gcn3
  %(2005).

\bibitem{grb051022A_ra85}
{Cameron}, P.~B. \emph{et~al.}
\newblock{GRB 051022: Radio counterpart}.
\newblock{\em GCN Circular} 4154, http://gcn.gsfc.nasa.gov/gcn3/4154.gcn3 (2005). 

\bibitem{grb050922C_z}
{D'Elia}, V. \emph{et~al.}
\newblock {GRB 050922C: UVES/VLT high resolution spectroscopy}.
\newblock {\em GCN Circular} 4044, http://gcn.gsfc.nasa.gov/gcn3/4044.gcn3
  (2005).

\bibitem{grb050922C_det}
{Golenetskii}, S. \emph{et~al.}
\newblock {Konus-Wind observation of GRB 050922C}.
\newblock {\em GCN Circular} 4030, http://gcn.gsfc.nasa.gov/gcn3/4030.gcn3
  (2005).

\bibitem{grb050922C_ra}
Frail, D.
\newblock {GRB 050922C data}.
\newblock {http://www.aoc.nrao.edu/$\sim$dfrail/grb050922c.dat}  (2005).

\bibitem{grb050904A_z}
{Kawai}, N. \emph{et~al.}
\newblock {GRB 050904: Subaru optical spectroscopy}.
\newblock {\em GCN Circular} 3937, http://gcn.gsfc.nasa.gov/gcn3/3937.gcn3
  (2005).

\bibitem{grb050904A_det}
{Cummings}, J. \emph{et~al.}
\newblock {GRB 050904: Swift-BAT detection of a probable burst}.
\newblock {\em GCN Circular} 3910, http://gcn.gsfc.nasa.gov/gcn3/3910.gcn3
  (2005).

\bibitem{grb050904A_ra}
Frail, D.~A.  \emph{et~al.}
\newblock {An energetic afterglow from a distant stellar explosion}.
\newblock {\em Astrophys. J. Lett.}{ \bf 646}, L99--L102 (2006).

\bibitem{grb050820A_z}
{Ledoux}, C. \emph{et~al.}
\newblock {VLT/UVES spectroscopy of GRB 050820}.
\newblock {\em GCN Circular} 3860, http://gcn.gsfc.nasa.gov/gcn3/3860.gcn3
  (2005).

\bibitem{grb050820A_det}
{Page}, M. \emph{et~al.}
\newblock {Subject: GRB 050820: Swift detection of a GRB}.
\newblock {\em GCN Circular} 3830, http://gcn.gsfc.nasa.gov/gcn3/3830.gcn3
  (2005).

\bibitem{grb050820A_ra}
{Cameron}, P.~B.  \emph{et~al.}
\newblock {GRB 050820A: Radio observation}.
\newblock {\em GCN Circular} 3847, http://gcn.gsfc.nasa.gov/gcn3/3847.gcn3
  (2005).

\bibitem{grb050730A_z}
{Holman}, M., {Garnavich}, P. \& {Stanek}, K.~Z.
\newblock {GRB 050730, spectra and optical photometry}.
\newblock {\em GCN Circular} 3716, http://gcn.gsfc.nasa.gov/gcn3/3716.gcn3
  (2005).

\bibitem{grb050730A_det}
{Holland}, S.~T. \emph{et~al.}
\newblock {GRB 050730: Swift-BAT detection of a weak burst}.
\newblock {\em GCN Circular} 3704, http://gcn.gsfc.nasa.gov/gcn3/3704.gcn3
  (2005).

\bibitem{grb050730A_ra}
{Cameron}, P.~B. \emph{et~al.}
\newblock {GRB 050730: Radio detection}.
\newblock {\em GCN Circular} 3761, http://gcn.gsfc.nasa.gov/gcn3/3761.gcn3
  (2005).

\bibitem{grb050724A_z}
{Prochaska}, J.~X. \emph{et~al.}
\newblock {GRB 050724: Secure host redshift from Keck}.
\newblock {\em GCN Circular} 3700, http://gcn.gsfc.nasa.gov/gcn3/3700.gcn3
  (2005).

\bibitem{grb050724A_det}
{Covino}, S. \emph{et~al.}
\newblock {GRB 050724: a short-burst detected by Swift}.
\newblock {\em GCN Circular} 3665, http://gcn.gsfc.nasa.gov/gcn3/3665.gcn3
  (2005).

\bibitem{grb050724A_ra}
 {Berger}, E. \emph{et~al.}
\newblock{The afterglow and elliptical host galaxy of the short {$\gamma$}-ray burst GRB 050724}.
\newblock {\em Nature} { \bf 438}, 988-990 (2005).

%\bibitem{grb050724A_ra}
%{Cameron}, P.~B.  \emph{et~al.}
%\newblock {GRB 050724: Radio Observation}.
%\newblock {\em GCN Circular} 3676. http://gcn.gsfc.nasa.gov/gcn3/3676.gcn3
  %(2005).

\bibitem{grb050603A_z}
{Berger}, E. \& {Becker}, G.
\newblock {GRB 050603: Redshift}.
\newblock {\em GCN Circular} 3520, http://gcn.gsfc.nasa.gov/gcn3/3520.gcn3
  (2005).

\bibitem{grb050603A_det}
{Golenetskii}, S. \emph{et~al.}
\newblock {Konus-Wind observation of GRB 050603}.
\newblock {\em GCN Circular} 3518, http://gcn.gsfc.nasa.gov/gcn3/3518.gcn3
  (2005).

\bibitem{grb050603A_ra}
{Cameron}, P.~B.   \emph{et~al.}
\newblock {GRB 050603: Radio detection}.
\newblock {\em GCN Circular} 3513, http://gcn.gsfc.nasa.gov/gcn3/3513.gcn3
  (2005).

\bibitem{grb050525A_z}
{Foley}, R.~J., {Chen}, H., {Bloom}, J. \& {Prochaska}, J.~X.
\newblock {GRB 050525a: Gemini/GMOS Spectra}.
\newblock {\em GCN Circular} 3483, http://gcn.gsfc.nasa.gov/gcn3/3483.gcn3
  (2005).

\bibitem{grb050525A_det}
{Band}, D. \emph{et~al.}
\newblock {GRB 050525: Swift detection of a bright, possibly short burst}.
\newblock {\em GCN Circular} 3466, http://gcn.gsfc.nasa.gov/gcn3/3466.gcn3
  (2005).

\bibitem{grb050525A_ra}
{Cameron}, P.~B.  \emph{et~al.}
\newblock {GRB 050525a: Radio observations}.
\newblock {\em GCN Circular} 3495, http://gcn.gsfc.nasa.gov/gcn3/3495.gcn3
  (2005).

\bibitem{grb050416A_z}
{Cenko}, S.~B. \emph{et~al.}
\newblock {GRB 050416(a): Host galaxy redshift determination}.
\newblock {\em GCN Circular} 3542, http://gcn.gsfc.nasa.gov/gcn3/3542.gcn3
  (2005).

\bibitem{grb050416A_det}
{Sakamoto}, T. \emph{et~al.}
\newblock {Swift BAT/XRT Detection of GRB 050416}.
\newblock {\em GCN Circular} 3264, http://gcn.gsfc.nasa.gov/gcn3/3264.gcn3
  (2005).

\bibitem{grb050416A_ra}
{Soderberg}, A.~M. \emph{et~al.}
\newblock {GRB 050416a: Radio detection}.
\newblock {\em GCN Circular} 3318, http://gcn.gsfc.nasa.gov/gcn3/3318.gcn3
  (2005).

\bibitem{grb050401A_z}
{Fynbo}, J.~P.~U. \emph{et~al.}
\newblock {GRB 050401: VLT spectroscopic redshift}.
\newblock {\em GCN Circular} 3176, http://gcn.gsfc.nasa.gov/gcn3/3176.gcn3
  (2005).

\bibitem{grb050401A_det}
{Golenetskii}, S. \emph{et~al.}
\newblock {Konus-Wind observation of GRB 050401}.
\newblock {\em GCN Circular} 3179, http://gcn.gsfc.nasa.gov/gcn3/3179.gcn3
  (2005).

\bibitem{grb050401A_ra}
{Soderberg}, A.~M.  \emph{et~al.}
\newblock {GRB 050401: Radio detection}.
\newblock {\em GCN Circular} 3187, http://gcn.gsfc.nasa.gov/gcn3/3187.gcn3
  (2005).

\bibitem{grb050315A_z}
{Kelson}, D. \& {Berger}, E.
\newblock {GRB 050315: Absorption redshift}.
\newblock {\em GCN Circular} 3101, http://gcn.gsfc.nasa.gov/gcn3/3101.gcn3
  (2005).

\bibitem{grb050315A_det}
{Parsons}, A. \emph{et~al.}
\newblock {Swift-BAT detection of GRB 050315}.
\newblock {\em GCN Circular} 3094, http://gcn.gsfc.nasa.gov/gcn3/3094.gcn3
  (2005).

\bibitem{grb050315A_ra}
{Soderberg}, A.~M.  \emph{et~al.}
\newblock {GRB 050315: Radio afterglow}.
\newblock {\em GCN Circular} 3102, http://gcn.gsfc.nasa.gov/gcn3/3102.gcn3
  (2005).

\bibitem{grb031203A_z}
{Prochaska}, J.~X. \emph{et~al.}
\newblock {GRB 031203: Magellan spectrum of possible GRB host}.
\newblock {\em GCN Circular} 2482, http://gcn.gsfc.nasa.gov/gcn3/2482.gcn3
  (2003).

\bibitem{grb031203A_ra}
{Soderberg}, A.~M. \emph{et~al.}
\newblock {The sub-energetic {$\gamma$}-ray burst GRB 031203 as a cosmic
  analogue to the nearby GRB 980425}.
\newblock {\em Nature}{ \bf 430}, 648--650  (2004).

\bibitem{grb031203A_det}
{Gotz}, D. \emph{et~al.}
\newblock {GRB 031203: A long GRB detected with INTEGRAL}.
\newblock {\em GCN Circular} 2459, http://gcn.gsfc.nasa.gov/gcn3/2459.gcn3
  (2003).

\bibitem{grb030329A_z}
{Caldwell}, N., {Garnavich}, P., {Holland}, S., {Matheson}, T. \& {Stanek},
  K.~Z.
\newblock {GRB 030329, optical spectroscopy}.
\newblock {\em GCN Circular} 2053, http://gcn.gsfc.nasa.gov/gcn3/2053.gcn3
  (2003).

\bibitem{grb030329A_det}
{Ricker}, G.~R. \emph{et~al.}
\newblock {GRB 030329}.
\newblock {\em IAU Circular} 8101,
  http://www.cfa.harvard.edu/iauc/08100/08101.html  (2003).

%\bibitem{grb030329A_ra}
%{Berger}, E. \emph{et~al.}
%\newblock {GRB 030329: Radio Observations}.
%\newblock {\em GCN Circular} 2014. http://gcn.gsfc.nasa.gov/gcn3/2014.gcn3
  %(2003).

\bibitem{grb030329A_ra}
{Berger}, E. \emph{et~al.}
\newblock{A common origin for cosmic explosions inferred from calorimetry of GRB030329}.
  \newblock{\em Nature} {\bf 426}, 154-157 (2003).


%\bibitem{grb030227A_z}
%Watson, D., Reeves, J.~N., Hjorth, J., Jakobsson, P. \& Pedersen, K.
%\newblock {Delayed soft X-ray emission lines in the afterglow of GRB 030227}.
%\newblock {\em Astrophys. J. Lett.}{ \bf 595}, L29--L32 (2003).

%\bibitem{grb030227A_det}
%{Gotz}, D. \emph{et~al.}
%\newblock {GRB 030227 localized by INTEGRAL}.
%\newblock {\em GCN Circular} 1895. http://gcn.gsfc.nasa.gov/gcn3/1895.gcn3
 % (2003).

%\bibitem{grb030227A_ra}
%Frail, D.
%\newblock {GRB 030227 data}.
%\newblock {http://www.aoc.nrao.edu/$\sim$dfrail/grb030227.dat},  (2003).

\bibitem{grb030226A_z}
{Greiner}, J.  \emph{et~al.}
\newblock {Two absorption systems in GRB 030226}.
\newblock {\em GCN Circular} 1886, http://gcn.gsfc.nasa.gov/gcn3/1886.gcn3
  (2003).

\bibitem{grb030226A_det}
{Suzuki}, M. \emph{et~al.}
\newblock {GRB 030226 (=U10893): A long GRB localized by the HETE WXM and SXC}.
\newblock {\em GCN Circular} 1888, http://gcn.gsfc.nasa.gov/gcn3/1888.gcn3
  (2003).

\bibitem{grb030226A_ra}
Frail, D.
\newblock {GRB 030226 data}.
\newblock {http://www.aoc.nrao.edu/$\sim$dfrail/grb030226.dat},  (2003).

\bibitem{grb021004A_z1}
Castro-Tirado, A.~J. {\em et al.} 
\newblock {GRB 021004: optical spectroscopy on Oct 11}.
\newblock {\em GCN Circular} 1635, http://gcn.gsfc.nasa.gov/gcn3/1635.gcn3  (2002).

\bibitem{grb021004A_z2}
{M{\"o}ller}, P. {\em et al.}
\newblock{Absorption systems in the spectrum of GRB 021004}.
\newblock{\em Astron. Astrophys.}  {\bf 396}, {L21-L24} (2002).


\bibitem{grb021004A_det}
{Shirasaki}, Y. \emph{et~al.}
\newblock {GRB 021004(HETE 2380): A long GRB localized by HETE in near-real
  time}.
\newblock {\em GCN Circular} 1565, http://gcn.gsfc.nasa.gov/gcn3/1565.gcn3
  (2002).

\bibitem{grb021004A_ra}
{Frail}, D.~A.  \emph{et~al.}
\newblock {GRB 021004: Radio observations}.
\newblock {\em GCN Circular} 1574, http://gcn.gsfc.nasa.gov/gcn3/1574.gcn3
  (2002).

\bibitem{grb021004A_ra15}
{Pooley}, G.
\newblock{GRB 021004 radio observation}.
\newblock{\em GCN Circular} 1604, http://gcn.gsfc.nasa.gov/gcn3/1604.gcn3 (2002).

\bibitem{grb021004A_ra486} 
{Berger}, E.  \emph{et~al.}
\newblock{GRB 021004 - VLA radio observations}.
 \newblock{\em GCN Circular} 1613, http://gcn.gsfc.nasa.gov/gcn3/1613.gcn3  (2002).

%\bibitem{grb020903A_z}
%{Soderberg}, A.~M. \emph{et~al.}
%\newblock {XRF 020903: Supernova}.
%\newblock {\em GCN Circular} 1554. http://gcn.gsfc.nasa.gov/gcn3/1554.gcn3
 % (2002).

%\bibitem{grb020903A_det}
%{Ricker}, G. \emph{et~al.}
%\newblock {GRB 020903(HETE 2314): An X-Ray Flash Localized by HETE}.
%\newblock {\em GCN Circular} 1530. http://gcn.gsfc.nasa.gov/gcn3/1530.gcn3
  %(2002).

%\bibitem{grb020903A_ra}
%{Berger}, E. \emph{et~al.}
%\newblock {Radio Observations of XRF 020903: The Missing Link}.
%\newblock {\em GCN Circular} 1555. http://gcn.gsfc.nasa.gov/gcn3/1555.gcn3
 % (2002).

\bibitem{grb020813A_z}
{Fiore}, F. \emph{et~al.}
\newblock {GRB 020813: High-resolution optical spectroscopy}.
\newblock {\em GCN Circular} 1524, http://gcn.gsfc.nasa.gov/gcn3/1524.gcn3
  (2002).

\bibitem{grb020813A_det}
{Villasenor}, J. \emph{et~al.}
\newblock {GRB 020813(HETE 2262): A long, bright burst localized by HETE in
  near-real time}.
\newblock {\em GCN Circular} 1471, http://gcn.gsfc.nasa.gov/gcn3/1471.gcn3
  (2002).

\bibitem{grb020813A_ra}
{Frail}, D.~A.  \emph{et~al.}
\newblock {GRB 020813, radio detection}.
\newblock {\em GCN Circular} 1490, http://gcn.gsfc.nasa.gov/gcn3/1490.gcn3
  (2002).

\bibitem{grb020813A_ra2}
Frail, D.
\newblock {GRB 020813 data}.
\newblock {http://www.aoc.nrao.edu/$\sim$dfrail/grb020813.dat}  (2002).

\bibitem{grb020405A_z}
{Masetti}, N. \emph{et~al.}
\newblock {GRB 020405: VLT spectroscopy}.
\newblock {\em GCN Circular} 1330, http://gcn.gsfc.nasa.gov/gcn3/1330.gcn3
  (2002).

\bibitem{grb020405A_det}
{Hurley}, K. \emph{et~al.}
\newblock {IPN localization of GRB 020405}.
\newblock {\em GCN Circular} 1325, http://gcn.gsfc.nasa.gov/gcn3/1325.gcn3
  (2002).

\bibitem{grb020405A_ra}
{Berger}, E. \emph{et~al.}
\newblock {GRB 020405, Radio observations}.
\newblock {\em GCN Circular} 1331, http://gcn.gsfc.nasa.gov/gcn3/1331.gcn3
  (2002).

\bibitem{grb011211A_z}
{Gladders}, M. \emph{et~al.}
\newblock {GRB 011211, spectrum and optical photometry}.
\newblock {\em GCN Circular} 1209, http://gcn.gsfc.nasa.gov/gcn3/1209.gcn3
  (2001).

\bibitem{grb011211A_det}
{Gandolfi}, G. et~al.
\newblock {BeppoSAX Alert: GRB 011211(=XRF 011211)}.
\newblock {\em GCN Circular} 1188, http://gcn.gsfc.nasa.gov/gcn3/1188.gcn3
  (2001).

\bibitem{grb011211A_ra}
Frail, D.
\newblock {GRB 011211 data}.
\newblock {http://www.aoc.nrao.edu/$\sim$dfrail/011211.dat}  (2001).

%\bibitem{grb011130A_z}
%{Jha}, S. \emph{et~al.}
%\newblock {GRB 011130: Magellan spectroscopy}.
%\newblock {\em GCN Circular} 1183. http://gcn.gsfc.nasa.gov/gcn3/1183.gcn3
  %(2001).

%\bibitem{grb011130A_det}
%{Ricker}, G. \emph{et~al.}
%\newblock {GRB 011130 (HETE 1864): An X-Ray Rich GRB Detected by HETE}.
%\newblock {\em GCN Circular} 1165. http://gcn.gsfc.nasa.gov/gcn3/1165.gcn3
  %(2001).

%\bibitem{grb011130A_ra}
%{Berger}, E.  \emph{et~al.}
%\newblock {GRB 011130, Radio Observations}.
%\newblock {\em GCN Circular} 1173. http://gcn.gsfc.nasa.gov/gcn3/1173.gcn3
 % (2001).

\bibitem{grb011121A_z}
Garnavich, P.~M.  \emph{et~al.}
\newblock {Discovery of the low-redshift optical afterglow of GRB 011121 and
  its progenitor supernova 2001ke}.
\newblock {\em Astrophys. J.}{ \bf 582}, 924--932 (2003).

\bibitem{grb011121A_det}
{Piro}, L. \emph{et~al.}
\newblock {BEPPOSAX GRB 011121}.
\newblock {\em GCN Circular} 1147, http://gcn.gsfc.nasa.gov/gcn3/1147.gcn3
  (2001).

%\bibitem{grb011121A_ra}
%{Subrahmanyan}, R.  \emph{et~al.}
%\newblock {GRB 011121: Radio Observations}.
%\newblock {\em GCN Circular} 1156. http://gcn.gsfc.nasa.gov/gcn3/1156.gcn3
 %(2001).

\bibitem{grb011121A_ra}
{Price}, P.~A. \emph{et~al.}
\newblock{GRB 011121: A massive star progenitor}.
\newblock{\em Astrophys. J. Lett.} {\bf  572}, {L51-L55} (2002).

\bibitem{grb010921A_z}
{Djorgovski}, S.~G. \emph{et~al.}
\newblock {GRB 010921: Spectroscopy of the host galaxy}.
\newblock {\em GCN Circular} 1108, http://gcn.gsfc.nasa.gov/gcn3/1108.gcn3
  (2001).

\bibitem{grb010921A_det}
{Ricker}, G. \emph{et~al.}
\newblock {HETE 1761: A bright GRB detected by HETE}.
\newblock {\em GCN Circular} 1096, http://gcn.gsfc.nasa.gov/gcn3/1096.gcn3
  (2001).

\bibitem{grb010921A_ra}
{Price}, P.~A. \emph{et~al.}
\newblock {GRB 010921: Discovery of the first high energy transient explorer
 afterglow}.
\newblock {\em Astrophys. J. Lett.}{ \bf 571}, L121--L125  (2002).

\bibitem{grb010222A_z}
{Jha}, S. \emph{et~al.}
\newblock {GRB 010222: Another absorption line system}.
\newblock {\em GCN Circular} 974, http://gcn.gsfc.nasa.gov/gcn3/974.gcn3  (2001).

\bibitem{grb010222A_det}
{Piro}, L.  \emph{et~al.}
\newblock {ALERT: GRB 010222: the brightest GRB observed by BeppoSAX}.
\newblock {\em GCN Circular} 959, http://gcn.gsfc.nasa.gov/gcn3/959.gcn3  (2001).

\bibitem{grb010222A_ra}
{Berger}, E. \emph{et~al.}
\newblock {GRB 010222, radio observations}.
\newblock {\em GCN Circular} 968, http://gcn.gsfc.nasa.gov/gcn3/968.gcn3  (2001).

\bibitem{grb010222A_ra2}
Frail, D.
\newblock {GRB 010222 data}.
\newblock {http://www.aoc.nrao.edu/$\sim$dfrail/010222.dat} (2001).


\bibitem{grb000926A_z}
{Fynbo}, J.~P.~U. \emph{et~al.}
\newblock {Spectroscopic redshift of GRB 000926}.
\newblock {\em GCN Circular} 807, http://gcn.gsfc.nasa.gov/gcn3/807.gcn3  (2000).

\bibitem{grb000926A_det}
{Hurley}, K.  \emph{et~al.}
\newblock {IPN triangulation of GRB 000926}.
\newblock {\em GCN Circular} 801, http://gcn.gsfc.nasa.gov/gcn3/801.gcn3  (2000).

%\bibitem{grb000926A_ra}
%{Frail}, D.~A.  \emph{et~al.}
%\newblock {GRB 000926, radio observations}.
%\newblock {\em GCN Circular} 805. http://gcn.gsfc.nasa.gov/gcn3/805.gcn3  (2000).

\bibitem{grb000926A_ra}
  {Harrison}, F.~A. \emph{et~al.}
\newblock{Broadband observations of the afterglow of GRB 000926: Observing the effect of inverse Compton scattering}.
\newblock{\em Astrophys. J. }{\bf 559}, {123-130} (2001).

\bibitem{grb000911A_z}
{Price}, P.~A. \emph{et~al.}
\newblock {The unusually long duration gamma-ray burst GRB 000911: Discovery of
  the afterglow and host galaxy}.
\newblock {\em Astrophys. J.}{ \bf 573}, 85--91 (2002).

\bibitem{grb000911A_det}
{Hurley}, K. \emph{et~al.}
\newblock {IPN triangulation of GRB 000911}.
\newblock {\em GCN Circular} 791, http://gcn.gsfc.nasa.gov/gcn3/791.gcn3  (2000).

%\bibitem{grb000911A_ra}
%{Berger}, E. \emph{et~al.}
%\newblock {GRB 000911, Radio and Optical observations}.
%\newblock {\em GCN Circular} 795. http://gcn.gsfc.nasa.gov/gcn3/795.gcn3  (2000).

\bibitem{grb000418A_z}
{Bloom}, J.~S., {Berger}, E., {Kulkarni}, S.~R., {Djorgovski}, S.~G. \&
  {Frail}, D.~A.
\newblock {The redshift determination of GRB 990506 and GRB 000418 with the
  echellete spectrograph imager on Keck}.
\newblock {\em Astron. J.}{ \bf 125}, 999--1005   (2003).

\bibitem{grb000418A_det}
{Hurley}, K. \emph{et~al.}
\newblock {IPN error box for GRB 000418}.
\newblock {\em GCN Circular} 642, http://gcn.gsfc.nasa.gov/gcn3/642.gcn3  (2000).

%\bibitem{grb000418A_ra}
%{Frail}, D.~A.  \emph{et~al.}
%\newblock {GRB 000418, Radio detection}.
%\newblock {\em GCN Circular} 655. http://gcn.gsfc.nasa.gov/gcn3/655.gcn3  (2000).

\bibitem{grb000418A_ra}
  {Berger}, E. \emph{et~al.}
\newblock{GRB 000418: A hidden jet revealed}.
\newblock {\em Astrophys. J.} {\bf 556}, {556-561} (2001).

\bibitem{grb000301C_z}
{Castro}, S.~M. \emph{et~al.}
\newblock {GRB 000301C: A precise redshift determination}.
\newblock {\em GCN Circular} 605, http://gcn.gsfc.nasa.gov/gcn3/605.gcn3  (2000).

\bibitem{grb000301C_det}
{Smith}, D.~A. \emph{et~al.}
\newblock {GRB 000301C: RXTE/ASM and IPN localizations}.
\newblock {\em GCN Circular} 568, http://gcn.gsfc.nasa.gov/gcn3/568.gcn3  (2000).

%\bibitem{grb000301C_ra250}
%{Bertoldi}, F.
%\newblock {GRB 000301C 250 GHz detection}.
%\newblock {\em GCN Circular} 580. http://gcn.gsfc.nasa.gov/gcn3/580.gcn3  (2000).

\bibitem{grb000301C_ra2}
{Berger}, E.  \emph{et~al.}
\newblock{A jet model for the afterglow emission from GRB 000301C}.
\newblock {\em Astrophys. J.}  {\bf  545},   {56-62} (2000)

%\bibitem{grb991216A_z1}
%{Vreeswijk}, P.~M. \emph{et~al.}
%\newblock {VLT spectra of GRB 991216}.
%\newblock {\em GCN Circular} 496. http://gcn.gsfc.nasa.gov/gcn3/496.gcn3  (1999).

\bibitem{grb991216A_z}
   {Vietri}, M., {Ghisellini}, G., {Lazzati}, D., {Fiore}, F. \& {Stella}, L.
\newblock {Illuminated, and enlightened, by GRB 991216}.
\newblock {\em Astrophys. J. Lett.}{ \bf 550}, L43-L46 (2001).

\bibitem{grb991216A_det}
{Kippen}, R.~M. \emph{et~al.}
\newblock {GRB 991216: Extremely bright BATSE burst}.
\newblock {\em GCN Circular} 463, http://gcn.gsfc.nasa.gov/gcn3/463.gcn3  (1999).

\bibitem{grb991216A_ra}
{Rol}, E. \emph{et~al.}
\newblock {GRB 991216 radio observations}.
\newblock {\em GCN Circular} 491, http://gcn.gsfc.nasa.gov/gcn3/491.gcn3  (1999).

\bibitem{grb991216A_ra2}
{Frail}, D.~A.   \emph{et~al.}
\newblock{The enigmatic radio afterglow of GRB 991216}.
\newblock{\em Astrophys. J. Lett. } {\bf 538}, {L129-L132} (2000).


\bibitem{grb991208A_z}
{Dodonov}, S.~N., {Afanasiev}, V.~L., {Sokolov}, V.~V., {Moiseev}, A.~V. \&
  {Castro-Tirado}, A.~J.
\newblock {GRB 991208 SAO-RAS spectroscopy}.
\newblock {\em GCN Circular} 475, http://gcn.gsfc.nasa.gov/gcn3/475.gcn3  (1999).

\bibitem{grb991208A_det}
{Hurley}, K. \emph{et~al.}
\newblock {IPN localization of GRB 991208}.
\newblock {\em GCN Circular} 450, http://gcn.gsfc.nasa.gov/gcn3/450.gcn3  (1999).

%\bibitem{grb991208A_ra}
%{Pooley}, G.
%\newblock {GRB 991208 15-GHz observations}.
%\newblock {\em GCN Circular} 457. http://gcn.gsfc.nasa.gov/gcn3/457.gcn3  (1999).

 \bibitem{grb991208A_ra}
{Galama}, T.~J. \emph{et~al.}
\newblock{The bright gamma-ray burst 991208: Tight constraints on afterglow models from observations of the early-time radio evolution}.
\newblock{\em Astrophys. J. Lett.}  {\bf 541}, {L45-L49} (2000).


%\bibitem{grb990510A_z}
%{Vreeswijk}, P.~M. \emph{et~al.}
%\newblock {VLT spectrum of GRB 990510}.
%\newblock {\em GCN Circular} 324. http://gcn.gsfc.nasa.gov/gcn3/324.gcn3  (1999).

\bibitem{grb990510A_ra}
Harrison, F.~A.  \emph{et~al.}
\newblock {Optical and radio observations of the afterglow from GRB990510:
  Evidence for a jet}.
\newblock {\em Astrophys. J. Lett.}{ \bf 523}, L121--L124 (1999).

\bibitem{grb990510A_det}
{Piro}, L.  \emph{et~al.}
\newblock {GRB 990510: Final BeppoSAX-WFC coordinates}.
\newblock {\em GCN Circular} 304, http://gcn.gsfc.nasa.gov/gcn3/304.gcn3  (1999).


\bibitem{grb990506A_det}
{Kippen}, R.~M.  \emph{et~al.}
\newblock {GRB 990506: BATSE Observations}.
\newblock {\em GCN Circular} 306, http://gcn.gsfc.nasa.gov/gcn3/306.gcn3  (1999).

\bibitem{grb990506A_ra}
{Taylor}, G.~B. \emph{et~al.}
\newblock {The rapidly fading afterglow from the gamma-ray burst of 1999 May
  6}.
\newblock {\em Astrophys. J. Lett.}{ \bf 537}, L17--L21 (2000).

\bibitem{grb990123A_z}
{Kelson}, D.~D., {Illingworth}, G.~D., {Franx}, M., {Magee}, D. \& {van
  Dokkum}, P.~G.
\newblock {GRB 990123}.
\newblock {\em IAU Circular} 7096,
  http://www.cfa.harvard.edu/iauc/07000/07096.html  (1999).

\bibitem{grb990123A_det}
{Kippen}, R.~M.  \emph{et~al.}
\newblock {GRB 990123: BATSE Observations}.
\newblock {\em GCN Circular} 224, http://gcn.gsfc.nasa.gov/gcn3/224.gcn3  (1999).

\bibitem{grb990123A_ra1}
{Frail}, D.~A.  \emph{et~al.}
\newblock {GRB 990123, new radio source}.
\newblock {\em GCN Circular} 211, http://gcn.gsfc.nasa.gov/gcn3/211.gcn3  (1999).

\bibitem{grb990123A_ra2}
 {Kulkarni}, S.~R. {\em et al.}
\newblock{Discovery of a radio flare from GRB 990123}.
\newblock {\em Astrophys. J. Lett.} {\bf 522},   {L97-L100}  (1999).

\bibitem{grb980703A_z}
{Djorgovski}, S.~G., {Kulkarni}, S.~R., {Goodrich}, R., {Frail}, D.~A. \&
  {Bloom}, J.~S.
\newblock {GRB 980703: Spectrum of the proposed optical counterpart}.
\newblock {\em GCN Circular} 137, http://gcn.gsfc.nasa.gov/gcn3/137.gcn3  (1998).

\bibitem{grb980703A_det}
{Levine}, A. \emph{et~al.}
\newblock {GRB 980703}.
\newblock {\em IAU Circular} 6966, 
  http://www.cfa.harvard.edu/iauc/06900/06966.html  (1998).

%\bibitem{grb980703A_ra}
%{Frail}, D.~A. \emph{et~al.}
%\newblock {GRB 980703: Radio transient}.
%\newblock {\em GCN Circular} 141. http://gcn.gsfc.nasa.gov/gcn3/141.gcn3  (1998).

\bibitem{grb980703A_ra}
{Frail}, D.~A. {\em et al.}
\newblock{The broadband afterglow of GRB 980703}.
 \newblock{\em Astrophys. J. Lett.} {\bf 590}, {992-998} (2003).
 

%\bibitem{grb980425A_z}
%{Tinney}, C., {Stathakis}, R., {Cannon}, R. \& {Galama}, T.
%\newblock {GRB 980425}.
%\newblock {\em IAU Circular} 6896.
 % http://www.cfa.harvard.edu/iauc/06800/06896.html,  (1998).

%\bibitem{grb980425A_det}
%{Kippen}, R.~M.  \emph{et~al.}
%\newblock {GRB 980425 BATSE observations}.
%\newblock {\em GCN Circular} 67. http://gcn.gsfc.nasa.gov/gcn3/067.gcn3  (1998).

%\bibitem{grb980425A_ra}
%{Wieringa}, M. \emph{et~al.}
%\newblock {GRB 980425: Radio Observations}.
%\newblock {\em GCN Circular} 63. http://gcn.gsfc.nasa.gov/gcn3/063.gcn3  (1998).

\bibitem{grb970828A_z}
{Djorgovski}, S.~G. \emph{et~al.}
\newblock {The afterglow and the host galaxy of the dark burst GRB 970828}.
\newblock {\em Astrophys. J.}{ \bf 562}, 654--663  (2001).

\bibitem{grb970828A_det}
{Remillard}, R., {Wood}, A., {Smith}, D. \& {Levine}, A.
\newblock {GRB 970828}.
\newblock {\em IAU Circular} 6726,
  http://www.cfa.harvard.edu/iauc/06700/06726.html  (1997).

%\bibitem{grb970828A_ra}
%{Frail}, D.~A. \& {Kulkarni}, S.~R.
%\newblock {GRB 970828}.
%\newblock {\em IAU Circular} 6730.
 % http://www.cfa.harvard.edu/iauc/06700/06730.html,  (1997).
    
\bibitem{grb970508A_z1}
{Metzger} \emph{et~al.}
\newblock {GRB 970508}.
\newblock {\em IAU Circular} 6655,
  http://www.cfa.harvard.edu/iauc/06600/06655.html  (1997).

\bibitem{grb970508A_z2}
 {Metzger}, M.~R. {\em et al.}
  \newblock{Spectral constraints on the redshift of the optical counterpart to the {$\gamma$}-ray burst of 8 May 1997}.
\newblock{\em Nature} {\bf 387}, {878-880} (1997).

\bibitem{grb970508A_z3}
  {Reichart}, D.~E.
 \newblock{The redshift of GRB 970508}.
\newblock{\em Astrophys. J. Lett.} {\bf 495},  {L99-L101} (1998).

\bibitem{grb970508A_det}
{Costa}, E. \emph{et~al.}
\newblock {GRB 970508}.
\newblock {\em IAU Circular} 6649,
  http://www.cfa.harvard.edu/iauc/06600/06649.html (1997).

%\bibitem{grb970508A_ra}
%{Frail}, D.~A. \emph{et~al.}
%\newblock {GRB 970508}.
%\newblock {\em IAU Circular} 6662.
 % http://www.cfa.harvard.edu/iauc/06600/06662.html,  (1997).

\bibitem{grb970508A_ra}
{Frail}, D.~A., {Waxman}, E. \& {Kulkarni}, S.~R.
  \newblock{A 450 day light curve of the radio afterglow of GRB 970508: Fireball calorimetry}.
 \newblock{\em Astrophys. J.} {\bf 537}, {191-204} (2000). 


\end{thebibliography}
\end{document}